\documentclass[aps,eqsecnum,showpacs,superscriptaddress]{revtex4}
\usepackage{graphicx,amssymb,amsbsy,bm,amsmath}

\usepackage{hyperref}

\newcommand{\intx}[1][]{\ensuremath{\int d^3 x{#1}}}
\newcommand{\intk}{\ensuremath{\int\frac{d^3 k}{(2\pi)^3}}}
\newcommand{\intq}{\ensuremath{\int\frac{d^3 q}{(2\pi)^3}}}
\newcommand{\intomega}{\ensuremath{\int^{+\infty}_{-\infty} d\omega}}
\newcommand{\bx}[1][]{\ensuremath{\mathbf{x}{#1}}}
\newcommand{\bk}{\ensuremath{\mathbf{k}}}
\newcommand{\bp}{\ensuremath{\mathbf{p}}}
\newcommand{\bq}{\ensuremath{\mathbf{q}}}
\newcommand{\ba}{\ensuremath{\mathbf{a}}}
\newcommand{\bhk}{\ensuremath{\widehat{\mathbf{k}}}}

\newcommand{\bhq}{\ensuremath{\widehat{\mathbf{q}}}}
\newcommand{\xt}[1][]{\ensuremath{\mathbf{x}{#1},t{#1}}}
\newcommand{\kt}{\ensuremath{\mathbf{k},t}}

\newcommand{\komega}{\ensuremath{k,\omega}}
\newcommand{\sbar}{\ensuremath{\bar{s}}}
\newcommand{\nn}{\nonumber}
\newcommand{\figskip}{\vspace{0.3in}}
\begin{document}
\title{Nonequilibrium relaxation in neutral BCS superconductors: \\
Ginzburg-Landau approach with Landau damping in real time}
\author{Saeed M. Alamoudi}
\email{alamoudi@kfupm.edu.sa} \affiliation{Department of Physics,
King Fahd University of Petroleum and Minerals, Dhahran, Saudi
Arabia}
\author{Daniel Boyanovsky}
\email{boyan@pitt.edu}\thanks{corresponding author.}
\affiliation{Department of Physics and Astronomy, University of
Pittsburgh, Pittsburgh, Pennsylvania 15260, USA}
\author{Shang-Yung Wang}
\email{swang@lanl.gov} \affiliation{Theoretical Division, MS B285,
Los Alamos National Laboratory, Los Alamos, New Mexico 87545, USA}
\date{July 2, 2002}

\begin{abstract}
We present a field-theoretical method to obtain consistently the
equations of motion for small amplitude fluctuations of the order
parameter directly in real time for a homogeneous, neutral BCS
superconductor. This method allows to study the nonequilibrium
relaxation of the order parameter as an initial value problem. We
obtain the Ward identities and the effective actions for small
phase  the amplitude fluctuations to one-loop order. Focusing on
the long-wavelength, low-frequency limit near the critical point,
we obtain the time-dependent Ginzburg-Landau effective action to
one-loop order, which is nonlocal as a consequence of Landau
damping. The nonequilibrium relaxation of the phase and amplitude
fluctuations is studied directly in real time. The long-wavelength
phase fluctuation (Bogoliubov-Anderson-Goldstone mode) is
\emph{overdamped} by Landau damping and the relaxation time scale
diverges at the critical point, revealing \emph{critical slowing
down}.

\end{abstract}

%74.20.Fg: BCS theory and its development
%74.40.+k: Fluctuations (noise, chaos, nonequilibrium superconductivity, localization, etc.)
%74.90.+n: Other topics in superconductivity

\preprint{LA-UR-02-2101}

\maketitle

\section{Introduction}

Nonequilibrium phenomena in superconductors continue to be the
focus of attention.  The dynamics of Josephson junctions, phase
slip phenomena in the dynamics of vortices and relaxation of the
order parameter and supercurrents are few examples of the
experimental effort that probe nonequilibrium aspects of
superconductivity.

Since the original work of Abrahams and Tsuneto \cite{tsuneto}
there has been an ongoing effort in trying to obtain the effective
time-dependent description of nonequilibrium phenomena from a
microscopic Bardeen-Cooper-Schrieffer (BCS) \cite{BCS,deGennes}
theory. Whereas the effective Ginzburg-Landau (GL) \cite{GL}
description in the \emph{static} limit was derived by Gor'kov
\cite{gorkov59,book:fetter}, the effective time dependent
Ginzburg-Landau description is still the focus of a substantial
theoretical effort. There is a large body of work that established
the validity of a time-dependent nonlinear Schroedinger equation
that describes the dynamics of the order parameter at zero
temperature \cite{fraser,wilczek,schakel,aitchison95,stone,palo}.

At \emph{finite} temperature the dynamical description is
complicated by the presence of Landau damping which prevents a
local description in time because the spectral densities feature
branch cuts that prevent a derivative expansion. This problem was
originally pointed out by Abrahams and Tsuneto \cite{tsuneto}. At
finite temperature Landau damping cuts are unavoidable and result
from processes that involve scatterings of quasiparticles in the
thermal bath. In derivations of the effective Lagrangian for
dynamical phenomena from a microscopic theory the Landau damping
contribution had often been ignored \cite{stoof}.

The absorptive contributions to the effective action of long
wavelength phase fluctuations at finite temperature have been
studied by Aitchison et al.\ \cite{aitchison97} for a neutral
BCS superconductor. These authors studied in detail the Landau
damping contributions to the effective action of phase
fluctuations and concluded that for temperatures $0<T<0.6\,T_c$
the effective propagator for the phase fluctuation
(Bogoliubov-Anderson-Goldstone mode) can be well approximated by
simple quasiparticle poles at complex energy and describe damped
excitations with a linear and temperature dependent dispersion
relation and narrow widths.

%new paragraph 1
An alternative approach to study nonequilibrium aspects of
superconductors is based on kinetic theory. Kopnin \cite{kopnin}
studied the nonequilibrium dynamics of flux flow in clean
superconductors but did not address the validity of the time
dependent Landau-Ginzburg description near the critical point.
Watts-Tobin et al.\ \cite{watts} studied the validity of
the Landau-Ginzburg description near the critical point for
\emph{dirty} superconductors where relaxational processes are
dominated by (elastic) collisions.

However, to the best of our knowledge, the description of the
relaxational dynamics for the amplitude and the phase of the order
parameter, as well as the validity of the Landau-Ginzburg
description \emph{near the critical point} in \emph{clean} neutral
superconductors had been the subject of several recent studies but
has not been completely understood. The region near the critical
temperature, where $|\Delta_0(T)| \ll T$ with $\Delta_0(T)$ the
finite-temperature gap (order parameter), is the region of
validity of the Ginzburg-Landau theory.

%end of new paragraph 1

%begin new paragraph 2

The interest on a deeper understanding of the time dependent
effective action for long-wavelength phase fluctuations has been
rekindled by several recent developments. Recently there has been
a substantial effort to obtain the time dependent effective action
of long-wavelength collective excitations associated with phase
fluctuations in $d$-wave superconductors
\cite{sharapov,takada,para,benfatto}.

In particular these studies
focused on the novel Carlson-Goldman modes \cite{carlson}, which
are Goldstone-like modes in \emph{charged} superconductors that
emerge near the critical temperature.
While at $T=0$ the Anderson-Higgs mechanism combines the Goldstone
and gauge fields into a gapped plasma mode, near the critical
temperature a novel quasiparticle Goldstone-like excitation, the
Carlson-Goldman mode, is present in charged superconductors. This
mode is a superposition of the Bogoliubov-Anderson-Goldstone mode,
present in \emph{neutral} superconductors and the long-range gauge
field which is screened at finite temperature. In
Ref.~\onlinecite{takada} it was pointed out that the existence of
this mode is associated both with screening and Landau damping of
phase fluctuations. The importance of the nonequilibrium dynamics
of long-wavelength \emph{phase} fluctuations has also been
highlighted recently within the context of high-temperature
superconductivity \cite{HTSC}.

Furthermore, recent experiments in ultracold alkali atoms have
demonstrated the trapping and cooling of \emph{fermionic} alkalis,
in particular ${}^{40}\mathrm{K}$ and ${}^6\mathrm{Li}$ \cite{schreck}. One goal of
this present experimental effort is to observe a transition to a
\emph{neutral fermi superfluid} for fermi systems with an
\emph{attractive} interaction between atoms in two different
hyperfine states \cite{stoof2}. Recently the spectrum of low energy
collective excitations in the \emph{collisionless} regime has been
studied, in particular focusing on the emergence of Goldstone or
phase fluctuations in these \emph{neutral Fermi
superfluids} \cite{mottelson}. A proposal for the detection of the
phase transition to a neutral Fermi superfluid in ${}^6\mathrm{Li}$ alkalis
relies on the spectrum of long-wavelength collective (Goldstone)
excitations \cite{zambelli}.

The interest on neutral BCS Fermi superfluids is
interdisciplinary, from the current experimental efforts in Fermi
alkalis and in mixtures with Bose alkalis \cite{schreck}, to
neutron superfluidity in nuclear matter and neutron stars. For a
recent discussion on neutral Fermi superfluids and their interest
in a wide variety of fields see Ref.~\onlinecite{pethick}.
Hence the study of the dynamics of
phase fluctuations in neutral Fermi superfluids (or neutral BCS
superconductors) continues to be of timely interest and of
experimental relevance.

%%%%end of new paragraph 2

\textbf{The goals of this article}: In this article we focus on
the nonequilibrium \emph{real-time} dynamics of phase and
amplitude fluctuations in neutral BCS superconductors in the
Ginzburg-Landau regime near the critical temperature. In
particular we  obtain the effective \emph{dynamical}
Ginzburg-Landau description of nonequilibrium relaxation of
long-wavelength, low-frequency fluctuations of the order parameter
\emph{near the critical point}.

While previous efforts, notably by Aitchison et al.\
\cite{aitchison95,aitchison97}, focused on the long-wavelength,
low-frequency effective action well below the critical temperature
for $0<T<0.6\,T_c$ our goal is to study the critical region
$|\Delta_0(T)| \ll T_c$ with $\Delta_0(T)$ the finite-temperature
gap. Our study is different from previous attempts in several
respects: (i) We implement the Schwinger-Keldysh formulation of
nonequilibrium field theory \cite{schwinger} along with the
recently introduced \emph{tadpole method} \cite{tadpole} to obtain
the equations of motion for small amplitude fluctuations of the
order parameter in \emph{real time}. (ii) The equations of motion
obtained with these methods are retarded, lead to the  Ward
identities and allow to establish the retarded effective action at
once. Furthermore, the equations of motion describe an initial
value problem that allows a real-time study of relaxation and
damping. (iii) We then focus on the Ginzburg-Landau regime
$|\Delta_0(T)|/T \ll 1$ and establish the \emph{dynamical}
Ginzburg-Landau effective action to one-loop order for
long-wavelength, low-frequency fluctuations. This effective action
is retarded and nonlocal because of Landau damping. (iv) We study
the time evolution of small phase and amplitude fluctuations from
the equilibrium configuration and reveal directly in real time the
effect of Landau damping.

\textbf{Main results}: Implementing the Schwinger-Keldysh
formulation of nonequilibrium field theory and the tadpole method
we obtain the retarded equations of motion for small fluctuations,
which in turn lead to Ward identities associated with (global)
gauge invariance both \emph{in} and \emph{out of} equilibrium.
From these equation of motion we obtain the retarded one-loop
effective action which is \emph{nonlocal} as a consequence of
Landau damping.

We then focus on the Ginzburg-Landau ($\Delta_0 \ll T$) region and
study the real-time relaxation of small phase and amplitude
fluctuations. While the spectral density for phase fluctuations
features a peak that suggests a Goldstone-like dispersion
relation, the relaxational dynamics is completely
\emph{overdamped} as a consequence of Landau damping.

Far away from the Ginzburg-Landau regime at low temperatures, the
spectral densities for both phase and amplitude fluctuations
feature narrow quasiparticle peaks confirming previous results
\cite{aitchison95,aitchison97}. In particular, the real-time
relaxation of long-wavelength phase fluctuations is weakly
\emph{underdamped} by Landau damping.

The article is organized as follows. In Sec.~\ref{sec:II} we
introduce the model and the linear response formulation to obtain
the equations of motion. In Sec.~\ref{sec:noneq} we introduce the
Schwinger-Keldysh formulation in the Nambu-Gor'kov formalism to
study the nonequilibrium aspects of Bogoliubov quasiparticles. In
Sec.~\ref{sec:eqnofmot} we introduce the tadpole method, obtain
the equations of motion directly in real time and cast them in
terms of an initial value problem. We obtain explicitly the
retarded self-energies to one-loop order and  their spectral
representations and obtain the one-loop retarded effective action.
In Sec.~\ref{sec:WI} we obtain the Ward identities and discuss the
static limit of the self-energies. In Sec.~\ref{sec:TDGL} we
obtain the effective time dependent Ginzburg-Landau description
focusing on the Ginzburg-Landau regime and the long-wavelength,
low-frequency limit. In this section we provide a thorough
numerical analysis of the real-time evolution of the relaxation of
phase and amplitude fluctuations. Section~\ref{sec:conclusions}
presents our conclusions and poses new directions. An appendix is
devoted to an alternative derivation of the Bogoliubov
transformation, which facilitates the Schwinger-Keldysh
nonequilibrium formulation.

\section{Preliminaries}\label{sec:II}
\subsection{Neutral BCS model}
The BCS Hamiltonian of a neutral electron gas is given by
\begin{eqnarray}
H&=&\sum_{\sigma=\uparrow,\downarrow}\intx\,\psi^\dagger_\sigma(\xt)
\left(-\frac{\nabla^2}{2m}\right)\psi_\sigma(\xt)
-g\intx\,\psi^\dagger_\uparrow(\xt)\psi^\dagger_\downarrow(\xt)
\psi_\downarrow(\xt) \psi_\uparrow(\xt),
\end{eqnarray}
where $\psi_{\sigma}(\xt)$ are the Heisenberg complex fields
representing electrons of mass $m$ and spin $\sigma$, and $g>0$ is
the strength of the $s$-wave pairing interaction between spin-up
and spin-down electrons close to the Fermi surface. In this
article, we set $\hbar=k_\mathrm{B}=1$. The field
$\psi_{\sigma}(\xt)$ and its Hermitian conjugate satisfy the
equal-time \emph{anticommutation} relations
\begin{gather}
\{\psi_\sigma(\xt),\psi^\dagger_{\sigma'}(\bx',t)\}=
\delta_{\sigma\sigma'}\delta^{(3)}(\bx-\bx'),\nn\\
\{\psi_\sigma(\xt),\psi_{\sigma'}(\bx',t)\}=
\{\psi^\dagger_\sigma(\xt),\psi^\dagger_{\sigma'}(\bx',t)\}=0.
\end{gather}
The Hamiltonian $H$ is invariant under the $U(1)$ gauge
transformation
\begin{gather}
\psi_\sigma(\xt)\to e^{i\theta}\psi_\sigma(\xt),\nn\\
\psi^\dagger_\sigma(\xt)\to e^{-i\theta}\psi^\dagger_\sigma(\xt),
\end{gather}
where $\theta$ is a constant phase. A consequence of this $U(1)$
gauge symmetry is conservation of the number of electrons. Indeed,
the number operator of electrons
\begin{equation}
N=\sum_{\sigma=\uparrow,\downarrow}\intx\,
\psi^\dagger_\sigma(\xt)\psi_\sigma(\xt)
\end{equation}
commutes with $H$ and hence is a constant of motion. However, it
is convenient to work in the grand-canonical ensemble in which the
grand-canonical Hamiltonian is given by
\begin{eqnarray}
K &\equiv& H-\mu N\nn\\
&=&\sum_{\sigma=\uparrow,\downarrow}\intx\,\psi^\dagger_\sigma(\xt)
\left(-\frac{\nabla^2}{2m}-\mu\right)\psi_\sigma(\xt)
-g\intx\,\psi^\dagger_\uparrow(\xt)\psi^\dagger_\downarrow(\xt)
\psi_\downarrow(\xt)\psi_\uparrow(\xt),\label{K}
\end{eqnarray}
where the chemical potential $\mu$ is the Lagrange multiplier
associated with conservation of number of electrons. The chemical
potential $\mu$ is determined by fixing the number of electrons
and in general is a function of the temperature. However, for the
situation under study in which the temperature is much lower than
the Fermi temperature, $\mu$ can be approximated by its
zero-temperature value, i.e., the Fermi energy. The Lagrangian
(density) corresponding to $K$ is given by
\begin{equation}
\mathcal{L}[\psi^\dagger,\psi]=\sum_{\sigma=\uparrow,\downarrow}
\psi^\dagger_\sigma\left(i\frac{\partial}{\partial t}+
\frac{\nabla^2}{2m}+\mu\right)\psi_\sigma + g
\psi_\uparrow^\dagger\psi^\dagger_\downarrow
\psi_\downarrow\psi_\uparrow.\label{L}
\end{equation}

Introducing the auxiliary complex scalar \emph{pair field} and its
Hermitian conjugate defined as
\begin{gather}
\Delta(\xt)=g\psi_\downarrow(\xt)\psi_\uparrow(\xt),\nn\\
\Delta^\dagger(\xt)=g\psi^\dagger_\uparrow(\xt)
\psi^\dagger_\downarrow(\xt), \label{Delta}
\end{gather}
and performing the Hubbard-Stratonovich
transformation\cite{negele}, the Lagrangian can be written as
\begin{equation}
\mathcal{L}[\psi^\dagger,\psi,\Delta^\dagger,\Delta]=
\sum_{\sigma=\uparrow,\downarrow}\, \psi^{\dagger}_{\sigma} \left(
i\frac{\partial}{\partial t}  + \frac{\nabla^2}{2m}  + \mu
\right)\psi_{\sigma} + \Delta^\dagger\psi_\downarrow\psi_\uparrow
+ \psi^\dagger_\uparrow\psi^\dagger_\downarrow \Delta-\frac{1}{g}
\Delta^\dagger\Delta. \label{Leff}
\end{equation}
We note that the pair field $\Delta(\xt)$ is \emph{not} a
dynamical field as there is no corresponding kinetic term in the
Lagrangian \eqref{Leff}.

In the superconducting phase, we decompose the paring field into
the condensate and noncondensate parts
\begin{equation}
\Delta(\xt)=\langle\Delta(\xt)\rangle+\chi(\xt),\quad\langle
\chi(\xt) \rangle =0,\label{decompose}
\end{equation}
where
$\langle\mathcal{O}(\xt)\rangle=\mathrm{Tr}[\rho\,\mathcal{O}(\xt)]/\mathrm{Tr}\rho$denotes
the \emph{expectation value} of the Heisenberg operator
$\mathcal{O}(\xt)$ in the \emph{initial density matrix} $\rho$,
$\langle\Delta(\xt)\rangle$ is the superconducting order
parameter, and $\chi(\xt)$ describes  the noncondensate operator.
The presence of the condensate $\langle\Delta(\xt)\rangle\ne 0$
leads to spontaneous breaking of the $U(1)$ gauge symmetry. In the
absence of explicit symmetry breaking external sources, the
condensate is homogeneous (i.e., space-time independent)
$\langle\Delta(\xt)\rangle=\Delta_0$, which is the situation under
consideration in this article.

\subsection{Real-time relaxation in linear response}

The goal of this article is to obtain \emph{directly in real time}
the equations of motion for small amplitude perturbations of the
homogeneous superconducting condensate in an initial value problem
formulation. Our strategy to study the relaxation of the
condensate perturbation as an initial value problem begins with
preparing a superconducting state slightly perturbed away from
equilibrium by applying an external source coupled to the pair
field. Once the external source is switched off, the perturbed
condensate must relax towards equilibrium. It is precisely this
\emph{real-time evolution} of the nonequilibrium fluctuations
around the condensate the focus of this article.

Let $\eta(\xt)$ be an external $c$-number source coupled to the
pair field $\Delta(\xt)$, then the Lagrangian given by
\eqref{Leff} becomes
\begin{equation}
\mathcal{L}[\psi^\dagger,\psi,\Delta^\dagger,\Delta]\to
\mathcal{L}[\psi^\dagger,\psi,\Delta^\dagger,\Delta]
+\Delta^\dagger\eta+\eta^\ast\Delta.\label{linearresponse}
\end{equation}
The presence of the external source $\eta$ will induce a (linear)
response of the system in the form of an induced expectation value
\begin{equation}
\langle \Delta(\xt) \rangle_\eta= \Delta_0+
\delta(\xt).\label{response}
\end{equation}
Here, $\langle\Delta(\xt)\rangle_{\eta}$ denotes the expectation
value of the paring field $\Delta(\xt)$ in the presence of the
external source, $\Delta_0=\langle \Delta(\xt) \rangle$ is the
homogeneous order parameter in the absence the external source,
and $\delta(\xt)$ is the space-time dependent perturbation of the
homogeneous condensate $\Delta_0$ \emph{induced by the external
source}. The linear response perturbation $\delta(\xt)$ vanishes
when the external source $\eta(\xt)$ vanishes at all times.  This
is tantamount to decomposing the field into the homogeneous
condensate ($\Delta_0$), a small amplitude perturbation induced by
the external source [$\delta(\xt)$], and the noncondensate part
[$\chi(\xt)$] as
\begin{equation}
\Delta(\xt)= \Delta_0+\delta(\xt)+\chi(\xt),\quad\langle \chi(\xt)
\rangle_\eta =0. \label{finshift}
\end{equation}

In linear response theory $\delta(\xt)$ can be expressed in terms
of the \emph{exact} retarded Green's function of the pair fields
in the absence of external source \cite{book:fetter,ivp}. An
experimentally relevant initial value problem formulation for the
real-time relaxation of the condensate perturbation can be
obtained by considering that the external source is adiabatically
switched on at $t= -\infty$ and switched off at $t=0$, i.e.,
\begin{equation}
\eta(\xt)=\eta(\bx)\,e^{\epsilon t}\,\Theta(-t), \quad\epsilon\to
0^+. \label{extsource}
\end{equation}
The adiabatic switching-on of the external source induces a
space-time dependent condensate perturbation $\delta(\xt)$, which
is prepared adiabatically by the external source with a given
value $\delta (\bx,0)$ at $t=0$ determined by $\eta(\bx)$. For
$t>0$ after the external source has been switched off, the
perturbed condensate will evolve in the absence of any external
source relaxing towards equilibrium. Thus, the external source
$\eta(\xt)$ is necessary for preparing an initial state at $t=0$
setting up an initial value problem. This method has been applied
to study a wide variety of relaxation phenomena in different
settings \cite{ivp,mikheev}, including the relaxation of
condensate fluctuations  in homogeneous Bose-Einstein condensates
\cite{boyanbec}.

Using the decomposition (\ref{finshift}) we expand the Lagrangian
density, and consistently with linear response, keep only the
linear terms in $\delta$ and $\delta^\ast$, which are the small
amplitude perturbations from the homogeneous condensate induced by
the external source $\eta$. The Lagrangian $\mathcal{L}$ becomes
(in the presence of the external source $\eta$)
\begin{equation}
\mathcal{L}[\psi^\dagger,\psi,\chi^\dagger,\chi]=
\mathcal{L}_0[\psi^\dagger,\psi,\chi^\dagger,\chi]
+\mathcal{L}_\mathrm{int}[\psi^\dagger,\psi,\chi^\dagger,\chi],\label{L1}
\end{equation}
with
\begin{eqnarray}
\mathcal{L}_0[\psi^\dagger,\psi,\chi^\dagger,\chi]&=&
\sum_{\sigma=\uparrow,\downarrow}\psi^{\dagger}_{\sigma} \left(
i\frac{\partial}{\partial t}  + \frac{\nabla^2}{2m}  + \mu
\right)\psi_{\sigma} + \Delta_0^\ast\psi_\downarrow\psi_\uparrow +
\psi^\dagger_\uparrow\psi^\dagger_\downarrow \Delta_0-\frac{1}{g}
\chi^\dagger\chi,\nn\\
\mathcal{L}_\mathrm{int}[\psi^\dagger,\psi,\chi^\dagger,\chi]&=&
(\delta^\ast+\chi^\dagger)\psi_\downarrow\psi_\uparrow-\frac{1}{g}
(\delta^\ast+\Delta_0^\ast)\chi+\eta^\ast\chi+\mathrm{H.c.},\label{Lint}
\end{eqnarray}
where we have discarded the $c$-number (field operators
independent) terms. We note that the Lagrangian
$\mathcal{L}[\psi^\dagger,\psi,\chi^\dagger,\chi]$ is obviously
invariant under the gauge transformations
\begin{eqnarray}
&&\psi_\sigma,\chi,\Delta_0,\delta,\eta \to e^{i\theta/2}
\psi_\sigma, e^{i\theta}\chi, e^{i\theta}
\Delta_0,e^{i\theta} \delta,  e^{i\theta}\eta, \nn \\
&&\psi^\dagger_\sigma,\chi^\dagger,\Delta_0^\ast,\delta^\ast,\eta^\ast
\to e^{-i\theta/2} \psi^\dagger_\sigma, e^{-i\theta}\chi^\dagger,
e^{-i\theta} \Delta_0,e^{-i\theta} \delta, e^{-i\theta}\eta,
\label{gauge}
\end{eqnarray}
which, as will be seen below, is at the heart of the Ward
identity.

Whereas in general a gauge transformation is invoked to fix the
condensate $\Delta_0$ to be \emph{real} for convenience, this
choice corresponds to \emph{fixing a particular gauge}, which in
turn hides the underlying gauge symmetry. In order to obtain the
Ward identity associated with this symmetry we will keep a complex
condensate $\Delta_0$ and analyze in detail the transformation
laws of the various contributions to the equations of motion.

The study of the static and dynamical properties of the BCS theory
is simplified by introducing the Nambu-Gor'kov formulation. Let us
introduce the two-component Nambu-Gor'kov fields
\cite{nambu,book:fetter}
\begin{equation}
\Psi(\xt)=
\begin{bmatrix}
\psi_\uparrow(\xt) \\
\psi^\dagger_\downarrow(\xt)
\end{bmatrix},\quad \Psi^\dagger(\xt)=
\left[\psi^\dagger_\uparrow(\xt),\psi_\downarrow(\xt)\right],
\label{spinorPsi}
\end{equation}
and the $2\times 2$ Pauli matrices
\begin{equation}
\sigma_+ = \begin{bmatrix}
0 & 1 \\
0 & 0
\end{bmatrix},\quad
\sigma_- = \begin{bmatrix}
0 & 0 \\
1 & 0
\end{bmatrix},\quad
\sigma_3 = \begin{bmatrix}
1 & 0 \\
0 & -1
\end{bmatrix}, \label{matrices}
\end{equation}
in terms of which the Lagrangian can be written as
\begin{equation}
\mathcal{L}[\Psi^\dagger,\Psi,\chi^\dagger,\chi]=
\mathcal{L}_0[\Psi^\dagger,\Psi,\chi^\dagger,\chi]+
\mathcal{L}_\mathrm{int}[\Psi^\dagger,\Psi,\chi^\dagger,\chi],
\end{equation}
with
\begin{eqnarray}
\mathcal{L}_0[\Psi^\dagger,\Psi,\chi^\dagger,\chi] &=&
\Psi^\dagger \bigg[i \frac{\partial}{\partial t}+\sigma_3\left(
\frac{\nabla^2}{2m}+\mu\right)+\sigma_+ \Delta_0 +
\sigma_-\Delta_0^\ast \bigg]\Psi -\frac{1}{g}\chi^\dagger\chi,\nn\\
\mathcal{L}_\mathrm{int}[\Psi^\dagger,\Psi,\chi^\dagger,\chi]&=&
\Psi^\dagger\sigma_+(\delta+\chi)\Psi+\frac{1}{g}
\chi^\dagger(\Delta_0+\delta)+ \chi^\dagger\eta+
\mathrm{H.c.}.\label{NGlag}
\end{eqnarray}

\section{Nonequilibrium formulation}\label{sec:noneq}
\subsection{Generating functional}
The general framework to study of nonequilibrium phenomena is the
Schwinger-Keldysh formulation \cite{schwinger}, which we briefly
review here in a manner that leads immediately to a path integral
formulation.

Consider that the system is described by an initial density matrix
$\rho$ and a perturbation is switched on at a time $t_0$, so that
the total Hamiltonian for $t>t_0$, $H(t)$, does not commute with
the initial density matrix. The expectation value of a Heisenberg
operator $\mathcal{O}(t)= U^{-1}(t,t_0)\mathcal{O}(t_0) U(t,t_0)$
is given by
\begin{equation}
\langle \mathcal{O}(t)\rangle = \frac{\mathrm{Tr}\rho
U^{-1}(t,t_0)\mathcal{O} U(t,t_0)}{\mathrm{Tr}\rho},
\label{expval}
\end{equation}
where $U(t,t_0)$ is the unitary time evolution operator in the
Heisenberg picture
\begin{equation}
U(t,t_0) = T \exp\left[-i \int_{t_0}^t dt'\,H(t')\right],
\label{evolop}
\end{equation}
with $T$ the time-ordering symbol. If the initial density matrix
$\rho$ describes a state in thermal equilibrium at inverse
temperature $\beta$ with the unperturbed Hamiltonian $H(t<t_0)=H$,
i.e.,
\begin{equation}
\rho = e^{-\beta H} = U(t_0-i\beta,t_0),
\end{equation}
then the expectation value (\ref{expval}) can be written in the
form
\begin{equation}
\langle \mathcal{O}(t)\rangle =
\frac{\mathrm{Tr}U(t_0-i\beta,t_0)U^{-1}(t_0,t)\mathcal{O}
U(t,t_0)}{\mathrm{Tr}U(t_0-i\beta,t_0)}. \label{expval2}
\end{equation}
The numerator of this expression has the following interpretation:
evolve in time from $t_0$ up to $t$, insert the operator
$\mathcal{O}$, evolve back from $t$ to the initial time $t_0$ and
down the imaginary axis in time from $t_0$ to $t_0-i\beta$. The
denominator describes the evolution in imaginary time which is the
familiar description of a thermal density matrix. We note that
unlike the $S$-matrix elements or transition amplitudes,
expectation values of Heisenberg operators require evolution
\emph{forward} and \emph{backward} in time (corresponding to the
$U$ and $U^{-1}$ on each side of the operator $\mathcal{O}$).

The time evolution operators have a path-integral representation
\cite{negele} in terms of the Lagrangian, and the insertion of
operators can be systematically handled by introducing sources
coupled linearly to the fields, i.e.,
\begin{equation}
\mathcal{L}[\Psi^\dagger,\Psi,\chi^\dagger,\chi]\to
\mathcal{L}[\Psi^\dagger,\Psi,\chi^\dagger,\chi]+\Psi^\dagger
J+J^\dagger \Psi+\chi^\dagger j+j^\ast \chi, \label{sources}
\end{equation}
where $J$ and $J^\dagger$ are Grassmann-valued variables. The
introduction of sources $J$, $J^\dagger$, $j$, and $j^\ast$  also
allows a systematic perturbative expansion. In such an expansion,
powers of operators are obtained by functional derivatives with
respect to these sources, which are set to zero after functional
differentiation. We note that the sources $j$, $j^\ast$ introduced
in \eqref{sources} to generate the perturbative expansion for the
pair fields in terms of functional derivatives with respect to
these, are \emph{different} from the external sources $\eta$,
$\eta^\ast$ introduced in (\ref{linearresponse}) to generate an
initial value problem in linear response and to displace the
condensate from equilibrium.

\begin{figure}[t]
\includegraphics[width=2.75in,keepaspectratio=true]{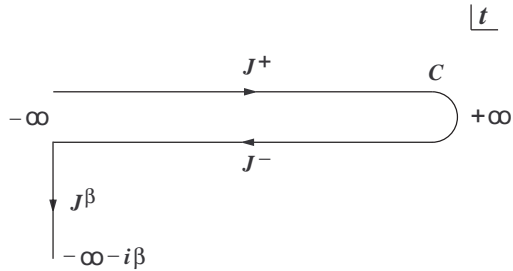}
\caption{The contour $\mathcal{C}$ in complex time plane in the
Schwinger-Keldysh formulation. It consists of a forward branch
running from $t=-\infty$ to $t=+\infty$, a backward branch from
$t=+\infty$ back to $t=-\infty$, and an imaginary branch from
$t=-\infty$ to $t=-\infty-i\beta$. The sources $J^\pm$ serve to
generate the nonequilibrium Green's functions of the Nambu-Gor'kov
fields.}\label{fig:ctpcontour}
\end{figure}

Since there are \emph{three} different time evolution operators,
the forward, backward and imaginary, we introduce \emph{three}
different sources for each one of these time evolution operators,
respectively. Taking $t_0 \to -\infty$, we are led to considering
the generating functional \cite{comment}
\begin{equation}
Z[J^+,J^-,J^\beta]=\mathrm{Tr}U(-\infty-i\beta,-\infty,J^\beta)
U(-\infty,+\infty,J^-)U(+\infty,-\infty;J^+),\label{genefunc}
\end{equation}
where $U(t_f,t_i;J)$ is the time evolution operator [see
\eqref{evolop}] in the presence of the source $J$ and for
simplicity of notation we have suppressed the spin index $\sigma$
and the complex conjugate of the sources $J^\dagger$. The
denominator in (\ref{expval2}) is given by $\mathrm{Tr}\rho =
Z[0,0,0]$. The generating functional $Z[J^+,J^-,J^\beta]$ can be
written as a path integral along the contour in complex time plane
(see Fig.~\ref{fig:ctpcontour})
\begin{equation}
Z[J^+,J^-,J^\beta] = \int
\mathcal{D}_\mathcal{C}\Psi^\dagger\mathcal{D}_\mathcal{C}\Psi\,
\exp\left[i\int_\mathcal{C}d^4x\,
\mathcal{L}_\mathcal{C}[\Psi^\dagger,\Psi;J]\right],\label{pathint}
\end{equation}
where
$\mathcal{D}_\mathcal{C}\Psi^\dagger\mathcal{D}_\mathcal{C}\Psi$
denotes the functional integration measure along the contour
$\mathcal{C}$ and
\begin{eqnarray}
\int_\mathcal{C}
d^4x\,\mathcal{L}_\mathcal{C}[\Psi^\dagger,\Psi;J] &=&
\int_{-\infty}^{+\infty}d^4x\,\mathcal{L}[\Psi^{\dagger +},
\Psi^+;J^+]-\int_{-\infty}^{+\infty}d^4x\,\mathcal{L}[\Psi^{\dagger
-},\Psi^-;J^-]\nn\\
&&+\,\int_{-\infty}^{-\infty-i\beta}d^4x\,
\mathcal{L}[\Psi^{\dagger\beta},\Psi^\beta;J^\beta],\label{pathL}
\end{eqnarray}
with
$\int_{-\infty}^{+\infty}d^4x\equiv\intx\int_{-\infty}^{+\infty}dt$,
etc. Because of the trace and the fermionic nature of the
operators, the path integral along the contour $\mathcal{C}$
requires \emph{antiperiodic boundary conditions} on the fields.
The superscripts $+$ and $-$ refer to fields defined in the upper
and lower branches, respectively, corresponding to forward ($+$)
and backward ($-$) time evolution, while the superscript $\beta$
refers to the field defined in the vertical branch running down
parallel to the imaginary axis. The negative sign in front of the
action along the backward branch is a result of the fact that
backward time evolution is determined by $U^{-1}(+\infty,-\infty)$
with $U$ the time evolution operator. The contour source $J$ that
enters in the contour Lagrangian $\mathcal{L}_\mathcal{C}$ in
(\ref{pathint}) takes the values of the sources $J^\pm$ and
$J^\beta$ in the respective branches as displayed in
Fig.~\ref{fig:ctpcontour}.

Functional derivatives with respect to the sources in the forward
branch give time-ordered Green's functions, those with respect to
the sources in the backward branch give the anti-time-ordered
Green's functions, and those with respect to the sources in the
imaginary branch give the usual imaginary-time (Matsubara) Green's
functions. While the sources $J^+$, $J^-$, and $J^\beta$
introduced to obtain the Green's functions via functional
differentiation are \emph{different} in the different branches, as
they generate the time-ordered, anti-time-ordered and Matsubara
Green's functions, respectively; the external source $\eta$, the
homogeneous condensate $\Delta_0$, and the departure from
equilibrium $\delta$ are $c$-numbers and hence are the \emph{same}
in all branches.

Writing the Lagrangian as a free and an interaction part as
$\mathcal{L}=\mathcal{L}_0+\mathcal{L}_\mathrm{int}$, the
generating functional can be written as a power series expansion
in the interaction part, which in turn can be generated by taking
functional derivatives with respect to the sources $J$,
$J^\dagger$ by identifying
\begin{gather}
\Psi^\pm \rightarrow \mp i\frac{\delta}{\delta
J^{\dagger\pm}},\quad
\Psi^{\dagger\pm} \rightarrow \pm i\frac{\delta}{\delta J^{\pm}},\nn\\
\Psi^\beta \to -i\frac{\delta}{\delta J^{\dagger\beta}},\quad
\Psi^{\dagger\beta} \to i\frac{\delta}{\delta
J^{\beta}}.\label{funcders}
\end{gather}
As a result, the full generating functional along the contour
$\mathcal{C}$ can be written as
\begin{equation}\label{PTgenfun}
Z[J] = \exp\left\{i\int_\mathcal{C}
d^4x\,\mathcal{L}_{\mathrm{int},\mathcal{C}}\left[-i\frac{\delta}{\delta
J^\dagger},i\frac{\delta}{\delta J}\right]\right\}Z_0[J],
\end{equation}
where free field generating functional $Z_0[J]$ is given by
(\ref{pathint}) and (\ref{pathL}) but with
$\mathcal{L}[\Psi^\dagger,\Psi;J]$ replaced by
$\mathcal{L}_0[\Psi^\dagger,\Psi;J]$.

\subsection{Green's functions}
The equation of motion for the \emph{free} Nambu-Gor'kov field
$\Psi$ in the presence of the source $J$ reads
\begin{equation}
\left[i \frac{\partial}{\partial t}+\sigma_3\left(
\frac{\nabla^2}{2m}+\mu\right)+\sigma_+ \Delta_0 +
\sigma_-\Delta_0^\ast\right]\Psi(\xt)= -J(\xt).\label{eqnofmot}
\end{equation}
The solution of this equation of motion is given by
\begin{equation}
\Psi_{J}(\xt) = -\int_\mathcal{C} d^4x'
S(\bx-\bx',t-t')J(\bx',t'), \label{sol}
\end{equation}
where $S(\bx-\bx',t-t')$ is the Green's function matrix along the
contour $\mathcal{C}$ and satisfies
\begin{equation}
\left[i \frac{\partial}{\partial t}+\sigma_3\left(
\frac{\nabla^2}{2m}+\mu\right)+\sigma_+ \Delta_0 +
\sigma_-\Delta_0^\ast\right]S(\bx-\bx',t-t')=
\delta_\mathcal{C}(t-t')\,\delta^{(3)}(\bx-\bx'),\label{green}
\end{equation}
with $\delta_\mathcal{C}(t-t')$ the Dirac delta function along the
contour $\int_\mathcal{C}dt'\,\delta_\mathcal{C}(t-t')=1$. The
Green's function $S(\bx-\bx',t-t')$ has the form
\begin{equation}
S(\bx-\bx',t-t')= S^>(\bx-\bx',t-t')\Theta_\mathcal{C}(t-t')+
S^<(\bx-\bx',t-t')\Theta_\mathcal{C}(t'-t), \label{Gcont}
\end{equation}
where $\Theta_\mathcal{C}(t-t')$ is the step function along the
contour and $S^\gtrless(\bx-\bx',t-t')$ obey the homogeneous
equations of motion.

The antiperiodic boundary conditions on the fields in the path
integral, a result of the trace over fermionic fields in
(\ref{genefunc}), lead to the following boundary condition on the
Green's function
\begin{equation}
\lim_{t_0\to -\infty} S(\bx-\bx',t_0-t')=-\lim_{t_0\to -\infty}
S(\bx-\bx',t_0-i\beta-t').\label{pbc}
\end{equation}
Since along the contour $t_0\to -\infty$ is the \emph{earliest}
time and $t_0-i\beta$ is therefore the \emph{latest} time,
\eqref{pbc} entails
\begin{equation}
\lim_{t_0\to -\infty}S^<(\bx-\bx',t_0-t')=-\lim_{t_0\to -\infty}
S^>(\bx-\bx',t_0-i\beta-t'),\quad \forall\;t',\label{pbccond}
\end{equation}
which is the Kubo-Martin-Schwinger (KMS) condition for equilibrium
correlation functions \cite{book:kadanoff}.

The free field generating functional $Z_0[J]$ is now obtained by
writing
\begin{gather}
\Psi(\bx,t) = \widetilde{\Psi}(\xt)+ \Psi_J(\xt),\nn\\
\Psi^\dagger(\bx,t) = \widetilde{\Psi}^\dagger(\xt)+
\Psi^\dagger_J(\xt),
\end{gather}
which leads to the result
\begin{equation}
Z_0[J] = Z_0[0]\exp\left[-i \int_\mathcal{C}d^4x\int_\mathcal{C}
d^4x' J^\dagger(x)S(x-x')J(x')\right], \label{genfun0}
\end{equation}
where and hereafter $x$ denotes the space-time coordinates $(\xt)$
for simplicity of notation. The source independent term $Z_0[0]$
will cancel between the numerator and the denominator in all
expectation values in (\ref{expval}).

Furthermore, we are interested in computing Green's functions of
\emph{finite real times} which are defined for fields in the
forward ($+$) and backward ($-$) time branches but not in the
imaginary branch. For these real-time Green's functions the
contributions to the generating functional from one source in the
imaginary branch and another source in either the forward or
backward branch vanish by the Riemann-Lebesgue lemma
\cite{tadpole,ivp}, since the time arguments are infinitely far
apart along the contour. Therefore the contour integrals of the
source terms and Green's functions in the generating functional
factorize into a term in which the sources are those either in the
forward and backward branches and another term in which
\emph{both} sources are in the imaginary branch
\cite{tadpole,ivp}. The latter term (with both sources in the
imaginary branch) cancel between numerator and denominator in
expectation values and the only remnant of the imaginary branch is
through the periodic boundary conditions along the full contour in
the Green's function.

Thus the generating functional for real-time Green's functions
simplifies to the following expression, defined solely along the
forward and backward time branches \cite{tadpole,ivp}
\begin{eqnarray}
Z[J^{\pm},J^{\dagger \pm}]&=& \exp\left[i\int_{-\infty}^{+\infty}
d^4 x\left\{\mathcal{L}_\mathrm{int}[-i\delta/\delta
J^{\dagger+},i\delta/\delta
J^+]-\mathcal{L}_\mathrm{int}[i\delta/\delta
J^{\dagger-},-i\delta/\delta J^-]\right\}\right]\nonumber\\
&&\times\,\exp\bigg\{-i\int_{-\infty}^{+\infty}d^4x
\int_{-\infty}^{+\infty}d^4x'
\big[J^{\dagger+}(x)S^{++}(x-x')J^+(x')+
J^{\dagger -}(x)\nn\\
&&\times\,S^{--}(x-x')J^-(x')-J^{\dagger
+}(x)S^{+-}(x-x')J^-(x')-J^{\dagger-}(x)S^{-+}(x-x')
J^+(x')\big]\bigg\},\label{realtimegen}
\end{eqnarray}
with
\begin{eqnarray}
&&S^{++}(x-x')=S^>(\bx-\bx',t-t')\Theta(t-t')+
S^<(\bx-\bx',t-t')\Theta(t'-t),\nn \\
&&S^{--}(x-x')=S^>(\bx-\bx',t-t')\Theta(t'-t)+
S^<(\bx-\bx',t-t')\Theta(t-t'),\nn\\
&&S^{-+}(x-x')=S^>(\bx-\bx',t-t'),\nn\\
&&S^{+-}(x-x')=S^<(\bx-\bx',t-t'),\label{gpm}
\end{eqnarray}
where now $-\infty \leq t, t' \leq +\infty$ and the superscripts
$+$, $-$ correspond to the sources defined on the forward ($+$)
and backward ($-$) time branches, respectively. An important issue
that must be highlighted at this stage, is that derivatives with
respect to sources in the forward ($+$) time branch correspond to
insertion of operators \emph{pre-multiplying} the density matrix
$\rho$ and derivatives with respect to sources in the backward
($-$) branch correspond to the insertion of operators
\emph{post-multiplying} the density matrix. That this is so is a
consequence of the fact that the density matrix evolves in time as
$U(t,t_0)\rho_0 U^{-1}(t,t_0)$ with $U(t,t_0)$ the time evolution
operator.

These four Green's functions are not independent because of the
identity
\begin{equation}
S^{++}(x-x')+S^{--}(x-x')-S^{+-}(x-x')-S^{-+}(x-x') = 0.
\label{identity}
\end{equation}
The diagonal elements in $S^{++}(x-x')$ are the normal Green's
functions, representing the propagation of single electrons,
whereas the off-diagonal elements are the anomalous Green's
functions, corresponding to the annihilation and creation of two
electrons of opposite spins, respectively.

The functions $S^\gtrless(x-x')$, which are solutions of the
homogeneous free field equation of motion, are simply related to
the correlation functions of the free Nambu-Gor'kov fields $\Psi$,
$\Psi^\dagger$. Indeed, taking variational derivatives of the free
field generating functional $Z_0[J]$ with respect to $J^{\pm}$,
$J^{\dagger\pm}$, one can show that
\begin{gather}
S^>_{ab}(x-x')= -i\langle \Psi_a(x)\Psi_b^\dagger(x')\rangle,\nn\\
S^<_{ab}(x-x')= i \langle
\Psi_b^\dagger(x')\Psi_a(x)\rangle,\label{gorkovpm}
\end{gather}
where and hereafter $a,b=1,2$ denote the Nambu-Gor'kov indices.
The expectation values in the expressions above are in the
non-interacting thermal density matrix which corresponds to the
quadratic part of the Lagrangian (Hamiltonian), i.e., the density
matrix that describes free Bogoliubov quasiparticles in thermal
equilibrium at inverse temperature $\beta$.

While the matrix elements $S^\gtrless_{ab}(x-x')$ can be obtained
through the usual Bogoliubov transformation to the quasiparticle
basis \cite{book:fetter}, we present in the Appendix an
alternative derivation of these correlation functions directly
from the spinor solutions of the homogeneous equations of motion.
We find that the correlation functions in (\ref{gorkovpm}) in the
infinite volume limit are given by
\begin{equation}
S^\gtrless_{ab}(x-x')= \int
\frac{d^3k}{(2\pi)^3}S^\gtrless_{ab}(k,t-t')\,e^{i\bk \cdot
(\bx-\bx')},
\end{equation}
where
\begin{gather}
S^>_{ab}(k,t-t')=
-i\left[[1-n_\mathrm{F}(E_k)]\,\mathcal{S}_{ab}(k) \,e^{-i
E_k(t-t')}+n_\mathrm{F}(E_k)\,\overline{\mathcal{S}}_{ab}(k)\,
e^{i E_k(t-t')}\right],\nn\\
S^<_{ab}(k,t-t') =  i\left[
n_\mathrm{F}(E_k)\,\mathcal{S}_{ab}(k)\,e^{-i E_k(t-t')} +
[1-n_\mathrm{F}(E_k)]\,\overline{\mathcal{S}}_{ab}(k)\,e^{i
E_k(t-t')}\right],\label{glesser}
\end{gather}
with
\begin{equation}
\mathcal{S}(k)=
\begin{bmatrix}
u^2_k & -u_k v_k^\ast \\
-u_k v_k & |v_k|^2
\end{bmatrix},\quad
\overline{\mathcal{S}}(k) =
\begin{bmatrix}
|v_k|^2 & u_k v_k^\ast \\
u_k v_k & u^2_k
\end{bmatrix}.\label{gmatrices}
\end{equation}
In the above expressions $k\equiv|\bk|$,
$n_\mathrm{F}(\omega)=1/(e^{\beta \omega}+1)$ is the Fermi-Dirac
distribution function, $u_k$ and $v_k$ satisfying
$u_k^2+|v_k|^2=1$ are the Bogoliubov coefficients [see
(\ref{coefs})], and $E_k$ is the energy of free Bogoliubov
quasiparticles (bogolons)
\begin{equation}
E_k=\sqrt{\xi_k^2+|\Delta_0|^2},\label{qpspectrum}
\end{equation}
where $\xi_k = k^2/2m-\mu$ (see the Appendix). Using the relation
$1-n_\mathrm{F}(\omega) = e^{\beta\omega}\,n_\mathrm{F}(\omega)$,
one can easily verify the KMS condition \cite{book:kadanoff}
\begin{equation}
S^>(k,t-i\beta-t') = -S^<(k, t-t').
\end{equation}
Hence, the correlation functions for the Nambu-Gor'kov fields
$\Psi$, $\Psi^\dagger$ that will enter in the nonequilibrium
perturbative expansion are completely determined by
(\ref{gorkovpm})-(\ref{qpspectrum}).

\subsection{Feynman rules}

From the generating functional of nonequilibrium Green's functions
(\ref{realtimegen}), it is clear that the effective interaction
Lagrangian relevant for the nonequilibrium calculations is given
by
\begin{equation}
\mathcal{L}_\mathrm{int}^\mathrm{eff}[\Psi^{\dagger
\pm},\Psi^\pm,\chi^{\dagger\pm},\chi^\pm]=\mathcal{L}_\mathrm{int}[\Psi^{\dagger
+},\Psi^+,\chi^{\dagger+},\chi^+]-\mathcal{L}_\mathrm{int}[\Psi^{\dagger-},
\Psi^-,\chi^{\dagger-},\chi^-],\label{efflag}
\end{equation}
where the fields $\Psi^{\dagger\pm}$, $\Psi^\pm$,
$\chi^{\dagger\pm}$ and $\chi^\pm$ are defined on the forward
($+$) and backwards ($-$) time branches respectively.
Consequently, this generating functional leads to the following
Feynman rules that define the perturbative expansion for
calculations of nonequilibrium expectation values.
\begin{itemize}
\item[(i)]{There are \emph{two} sets of interaction vertices defined by
$\mathcal{L}_\mathrm{int}^\mathrm{eff}
[\Psi^{\dagger\pm},\Psi^\pm,\chi^{\dagger\pm},\chi^\pm]$: those in
which the fields are in the forward ($+$) branch and those in
which the fields are in the backward ($-$) branch. There is a
relative minus sign between these two types of vertices. }
\item[(ii)]{For each kind of fields there are \emph{four} types of
Green's functions corresponding to correlations between fields
defined in the forward or backward branch.

The Green's functions of the Nambu-Gor'kov fields $\Psi$,
$\Psi^\dagger$ are given by (\ref{gpm}) in terms of the
correlation functions displayed in (\ref{gorkovpm}) that are
completely determined by (\ref{glesser}), (\ref{gmatrices}) and
(\ref{qpspectrum}).

The Green's functions of the pair fields $\chi$, $\chi^\dagger$
can be obtained in an analogous manner. However, due to the
nonpropagating nature of the pair fields, the Green's functions
are local in time and hence those of fields defined in different
branches vanish identically.}
\item[(iii)]{The combinatoric factors, trace over the Nambu-Gor'kov
indices for fermion loops, etc., are the same as in the
imaginary-time (Matsubara) formulation.}
\end{itemize}

\section{Relaxation of condensate perturbations: An initial value problem}\label{sec:eqnofmot}
\subsection{Equations of motion}
The equations of motion for the small amplitude superconducting
condensate perturbation $\delta(x)$ induced by the external source
$\eta(x)$ is obtained by implementing the \emph{tadpole method}
\cite{tadpole,ivp}. This method begins by writing the pair fields
$\Delta^\pm(x)$ in the forward ($+$) and backward ($-$) time
branches as
\begin{equation}
\Delta^\pm(x)= \Delta_0+\delta(x)+\chi^\pm(x),
\end{equation}
where $\Delta_0$ is the homogeneous condensate in the absence of
external source, $\delta(x)$ is the perturbation of the condensate
induced by the external source which vanishes in the absence of
external source, and $\chi^\pm(x)$ are the noncondensate part of
the pair fields in the forward and backward time branches. The
external source $\eta(x)$ are $c$-number fields and hence taken
the same value in both forward and backward branches. The strategy
to obtain the equations of motion for small amplitude condensate
perturbation $\delta(x)$, $\delta^\ast(x)$ is to consider the
interaction Lagrangian
$\mathcal{L}_\mathrm{int}[\Psi^\dagger,\Psi,\chi^\dagger,\chi]$
given in \eqref{NGlag} in perturbation theory and impose the
\emph{tadpole condition}
\begin{equation}
\langle \chi^{\pm}(x) \rangle_\eta = \langle \chi^{\dagger\pm}(x)
\rangle_\eta =0 \label{tadpolecond}
\end{equation}
order by order in the perturbative expansion, but, consistent with
linear response, only keep contributions linear in $\delta(x)$,
$\delta^\ast(x)$.

\begin{figure}[t]
\includegraphics[width=3.0in,keepaspectratio=true]{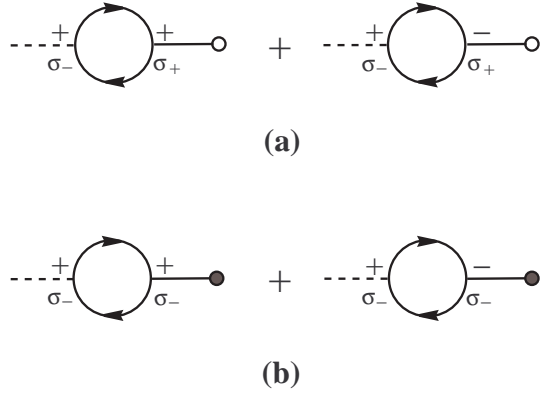}
\caption{Feynman diagrams contributing to the retarded
self-energies (a) $\Sigma_{11}(x-x')$ and (b) $\Sigma_{12}(x-x')$
to one-loop order. A solid line with arrow denotes the
Nambu-Gor'kov field, a dashed line denotes the pair field, and a
solid line with open (filled) circle denotes the condensate
perturbation $\delta$ ($\delta^\ast$).}\label{fig:sigma}
\end{figure}

Using the Feynman rules described in the previous section, to
one-loop order the tadpole condition
$\langle\chi^+(0)\rangle_\eta=\langle\chi^{\dagger+}(0)\rangle_\eta=0$
leads to the following expression
\begin{gather}
\int d^4x\langle\chi^+(0)\chi^{\dagger+}(x)
\rangle\bigg[\frac{\delta(x)}{g}+\int d^4x'
\big[\Sigma_{11}(x-x')\delta(x')+ \Sigma_{12}(x-x')
\delta^\ast(x')\big]+\mathcal{T}(\Delta_0,\Delta^\ast_0)-\eta(x)\bigg]=0,\nn\\
\int d^4x\langle\chi^{\dagger+}(0)\chi^+(x)
\rangle\bigg[\frac{\delta^\ast(x)}{g}+\int d^4x'
\big[\Sigma_{22}(x-x')\delta^\ast(x')+ \Sigma_{21}(x-x')
\delta(x')\big]+\mathcal{T}^\ast(\Delta_0,\Delta^\ast_0)-\eta^\ast(x)\bigg]=0,
\end{gather}
where $\Sigma_{ab}(x-x')$ are the retarded self-energies of the
pair fields and the \emph{tadpole}
$\mathcal{T}(\Delta_0,\Delta^\ast_0)$ denotes terms independent of
the condensate perturbation $\delta$, $\delta^\ast$. The diagrams
for $\Sigma_{11}(x-x')$ and $\Sigma_{12}(x-x')$ to one-loop order
are depicted in Fig.~\ref{fig:sigma}, and those for
$\Sigma_{22}(x-x')$ and $\Sigma_{21}(x-x')$ can be obtained in an
analogous manner. Explicitly to one-loop order we find
\begin{eqnarray}
\Sigma_{11}(x-x') & =&
-i\left[\langle\Psi^{\dagger+}(x)\sigma_-\Psi^+(x)\Psi^{\dagger+}(x')
\sigma_+\Psi^+(x')
\rangle-\langle\Psi^{\dagger+}(x)\sigma_-\Psi^+(x)\Psi^{\dagger-}(x')
\sigma_+\Psi^-(x')\rangle\right]\nn \\
&=& -i\,\textrm{tr}\left[\sigma_- S^{++}(x-x')\sigma_+
S^{++}(x'-x)-\sigma_- S^{+-}(x-x')\sigma_+ S^{-+}(x'-x)\right],\nn\\
\Sigma_{12}(x-x')&=&
-i\left[\langle\Psi^{\dagger+}(x)\sigma_-\Psi^+(x)\Psi^{\dagger+}(x')
\sigma_-\Psi^+(x')
\rangle-\langle\Psi^{\dagger+}(x)\sigma_-\Psi^+(x)\Psi^{\dagger-}(x')
\sigma_-\Psi^-(x')\rangle\right] \nn \\
&=& -i\,\textrm{tr}\left[\sigma_- S^{++}(x-x')\sigma_-
S^{++}(x'-x) -\sigma_-S^{+-}(x-x')\sigma_- S^{-+}(x'-x)\right],
\label{sigma1112}
\end{eqnarray}
where $\mathrm{tr}$ denotes the trace over the Nambu-Gor'kov
indices. The expressions for $\Sigma_{22}(x-x')$ and
$\Sigma_{21}(x-x')$ can be obtained, respectively, from
expressions in (\ref{sigma1112}) through the replacement $\sigma_+
\leftrightarrow \sigma_-$.

\begin{figure}[t]
\includegraphics[width=2.75in,keepaspectratio=true]{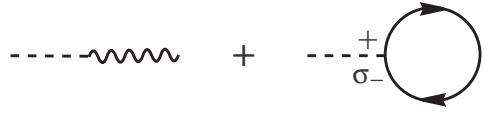}
\caption{Feynman diagrams contributing to the tadpole
$\mathcal{T}(\Delta_0,\Delta^\ast_0)$ to one-loop order. A wiggly
line denotes the homogeneous condensate $\Delta_0$.}
\label{fig:tadpole}
\end{figure}

The diagrams for the tadpole $\mathcal{T}(\Delta_0,\Delta^\ast_0)$
are depicted in Fig.~\ref{fig:tadpole}, from which we find
\begin{eqnarray}
\mathcal{T}(\Delta_0,\Delta^\ast_0)& = &\frac{\Delta_0}{g}-
\langle \Psi^{\dagger+}(x)\sigma_-\Psi^+(x)\rangle\nn\\
&=& \frac{\Delta_0}{g}+i\,\mathrm{tr}\left[\sigma_- S^<(0)\right].
\label{tadpole1}
\end{eqnarray}
A simple calculation with the nonequilibrium Green's functions of
the Nambu-Gor'kov fields obtained in the previous section shows
that the tadpole to one-loop order is given by
\begin{equation}
\mathcal{T}(\Delta_0,\Delta^\ast_0)=\Delta_0\left[\frac{1}{g}-\intq
\frac{1-2n_\mathrm{F}(E_q)}{2 E_q}\right],\label{tadpoleterm}
\end{equation}
where $E_q$ on the right-hand side of \eqref{tadpoleterm} is
understood to be a function of $|\Delta_0|$ [see
\eqref{qpspectrum}] and, hereafter, the momentum integral $\intq$
is restricted to the electronic states near the Fermi surface.

It is customary to rewrite the integral over the momentum $\bq$ as
being over the energy $\xi$ (measured from the Fermi surface) at
the expense of introducing the density of states
$\mathcal{N}(\xi)$, which is taken to be constant near the Fermi
surface, and cutting the integral off at $\pm\omega_\mathrm{D}$
with $\omega_\mathrm{D}$ being the Debye energy, thus leading to
\begin{equation}
\mathcal{T}(\Delta_0,\Delta^\ast_0)=\Delta_0\left[\frac{1}{g}-
\mathcal{N}(0)\int_0^{\omega_\mathrm{D}}\frac{d\xi}{E}
\,[1-2n_\mathrm{F}(E)]\right], \label{tadpoleterm2}
\end{equation}
where $E=\sqrt{\xi^2+|\Delta_0|^2}$ and $\mathcal{N}(0)=m
k_\mathrm{F}/2\pi^2$ is the density of states at the Fermi
surface. Setting the external source $\eta=0$, the equilibrium
condition for the homogeneous condensate
$\mathcal{T}(\Delta_0,\Delta^\ast_0)=0$ becomes
\begin{equation}
\Delta_0\left[\frac{1}{g}-
\mathcal{N}(0)\int_0^{\omega_\mathrm{D}}\frac{d\xi}{E}
\,[1-2n_\mathrm{F}(E)]\right]=0. \label{equicond}
\end{equation}
In equilibrium and below the critical temperature, the condensate
$\Delta_0\neq 0$. Therefore (\ref{equicond}) leads to the
finite-temperature BCS gap equation that determines $\Delta_0(T)$:
\begin{equation}
g\mathcal{N}(0)\int_0^{\omega_\mathrm{D}}\frac{d\xi}{E}
\,[1-2n_\mathrm{F}(E)]=1.\label{gapeq}
\end{equation}

Introducing the space Fourier transforms for $\delta(x)$,
$\eta(x)$ and $\Sigma_{ab}(x-x')$ as
\begin{equation*}
\delta(\xt)=\intk\,\delta_\bk(t)\,e^{i\bk\cdot\bx},
\end{equation*}
etc., we find the equations of motion in momentum space to be
given by
\begin{gather}
\delta_\bk(t)+ g\int_{-\infty}^{+\infty}
dt'\left[\Sigma_{11}(\bk,t-t')\,\delta_\bk(t')+
\Sigma_{12}(\bk,t-t')\,\delta^\ast_{-\bk}(t')
\right]+g\mathcal{T}(\Delta_0,\Delta^\ast_0)\,
\delta^{(3)}(\bk)-g\eta_\bk(t)=0,\nn\\
\delta^\ast_{-\bk}(t)+g\int_{-\infty}^{+\infty}
dt'\left[\Sigma_{22}(\bk,t-t')\,\delta^\ast_{-\bk}(t')+
\Sigma_{21}(\bk,t-t')\,\delta_\bk(t')\right]+
g\mathcal{T}^\ast(\Delta_0,\Delta^\ast_0)\,
\delta^{(3)}(\bk)-g\eta^\ast_{-\bk}(t)=0.\label{eom1}
\end{gather}
The equations of motion obtained from the tadpole condition
$\langle \chi^-(0)\rangle_\eta=\langle
\chi^{\dagger-}(0)\rangle_\eta=0$ are the same as those given in
the above expressions. While the equations of motion \eqref{eom1}
are obtained to one-loop order, it is straightforward to conclude
after a simple diagrammatic analysis that the structure of the
equations of motion obtained above is general and valid to all
orders in perturbation theory.

From the explicit expressions for the self-energies or by taking
complex conjugation of the equation of motion for $\delta_\bk(t)$,
one can show that
\begin{gather}
\Sigma_{21}(\kt-t')=\Sigma^\ast_{12}(-\kt-t'),\nn\\
\Sigma_{22}(\kt-t')=\Sigma^\ast_{11}(-\kt-t'). \label{conjuga}
\end{gather}
Furthermore, rotational and parity invariance imply that the
self-energies are only functions of $k$. Upon expressing the
nonequilibrium Green's functions in terms of the correlation
functions $S^\gtrless$, one finds immediately that the retarded
self-energies have the following causal structure
\begin{gather}
\Sigma_{11}(\bk,t-t')=\Sigma^R_{11}(k,t-t')\,\Theta(t-t'),\nn\\
\Sigma_{12}(\bk,t-t')=\Sigma^R_{12}(k,t-t')\,\Theta(t-t').
\end{gather}
Before proceeding further, we note that the invariance of the
Lagrangian under the global gauge transformation (\ref{gauge})
implies that \emph{all} the terms in the equations of motion for
$\delta$ ($\delta^\ast$) must transform as $\delta$
($\delta^\ast$) itself under the gauge transformation. This in
turn entails that the normal self-energies $\Sigma_{11}$ and
$\Sigma_{22}$ must be invariant under the gauge transformation
(\ref{gauge}), while the anomalous ones $\Sigma_{12}$ and
$\Sigma_{21}$ must transform as $\Delta^2_0$ and $\Delta^{\ast
2}_0$, respectively. Thus, it is proves convenient to write
\begin{gather}
\Sigma^R_{11}(k,t-t')=\Sigma^{R\ast}_{22}(k,t-t')=\Sigma_D(k,t-t'),\nn\\
\Sigma^R_{12}(k,t-t')=\Sigma^{R\ast}_{21}(k,t-t')=
(\Delta_0/\Delta_0^\ast)\Sigma_O(k,t-t'),\label{sigmaDO}
\end{gather}
where both $\Sigma_D$ and$\Sigma_O$ are \emph{invariant} under the
gauge transformation \eqref{gauge}. While rewriting the
self-energies in this manner may seem a redundant exercise, the
main point is to highlight and make explicit their transformation
laws under the gauge transformation. This is an important aspect
that needs to be addressed carefully in order to extract the Ward
identities, an \emph{exact} result of the underlying gauge
symmetry to be explored below.

The gauge invariant self-energies $\Sigma_D$ and $\Sigma_O$ can be
written in terms of their spectral representation as
\begin{eqnarray}
&&\Sigma_D(k,t-t')=\intomega\left[\overline{A}(\komega)\sin\omega(t-t')
+i\overline{S}(\komega)\cos\omega(t-t')\right],\nn\\
&&\Sigma_O(k,t-t')=\intomega\,\widehat{A}(\komega)\sin\omega(t-t').\label{sigmas}
\end{eqnarray}
The symmetric ($\overline{S}$) and antisymmetric ($\overline{A}$,
$\widehat{A}$) spectral functions are, respectively, even and odd
functions of $\omega$ and are  to be given by
\begin{eqnarray}
\overline{S}(\komega)&=&-\frac{1}{4}\intq
\bigg\{[1-n_\mathrm{F}(E_q)-n_\mathrm{F}(E_p)]
\left(\frac{\xi_q}{E_q} +
\frac{\xi_p}{E_p}\right)[\delta(\omega-E_q-E_p)+
\delta(\omega+E_q+E_p)]\nn\\&&
-\,[n_\mathrm{F}(E_q)-n_\mathrm{F}(E_p)]\left(\frac{\xi_q}{E_q}-
\frac{\xi_p}{E_p}\right)[\delta(\omega-E_q+E_p)+
\delta(\omega+E_q-E_p)]\bigg\},\nn\\
\overline{A}(\komega)&=&-\frac{1}{4}\intq
\bigg\{[1-n_\mathrm{F}(E_q)-n_\mathrm{F}(E_p)]\left(1+
\frac{\xi_q\xi_p}{E_qE_p}\right)
[\delta(\omega-E_q-E_p)-\delta(\omega+E_q+E_p)]\nn\\&&
-\,[n_\mathrm{F}(E_q)-n_\mathrm{F}(E_p)]\left(1-\frac{\xi_q\xi_p}{E_qE_p}\right)
[\delta(\omega-E_q+E_p)-\delta(\omega+E_q-E_p)]\bigg\},\nn\\
\widehat{A}(\komega)&=&\frac{1}{4}\intq\frac{|\Delta_0|^2}{E_qE_p}
\Big\{[1-n_\mathrm{F}(E_q)-n_\mathrm{F}(E_p)][\delta(\omega-E_q-E_p)-
\delta(\omega+E_q+E_p)]\nn\\
&&+\,[n_\mathrm{F}(E_q)-n_\mathrm{F}(E_p)][\delta(\omega-E_q+E_p)
-\delta(\omega+E_q-E_p)]\Big\},\label{spectral}
\end{eqnarray}
where $p=|\bk+\bq|$ and $\xi_q=q^2/2m-\mu$.

The terms proportional to
$[1-n_\mathrm{F}(E_q)-n_\mathrm{F}(E_p)]$ in the above expressions
correspond to processes in which two quasiparticles are either
created or destroyed, whereas the terms proportional to
$[n_\mathrm{F}(E_q)-n_\mathrm{F}(E_p)]$ arise from \emph{Landau
damping} processes corresponding to scattering of quasiparticle
excitations in the medium.

In an experimental situation the dynamical evolution of the small
amplitude condensate perturbation is studied by preparing a
superconducting state slightly perturbed away from equilibrium by
adiabatically coupling to some external source in the infinite
past. Once the source is switched-off at time $t=0$ the perturbed
condensate relaxes towards equilibrium and the relaxation dynamics
is studied. As discussed above, this experimental situation can be
realized within the real-time formulation described here by taking
the spatial Fourier transform of the external source to be of the
form
\begin{equation}
\eta_\bk(t)=\eta_\bk\,e^{\epsilon t}\,\Theta(-t),\quad \epsilon
\to 0^+.\label{adsource}
\end{equation}
The $\epsilon$-term serves to switch on the source adiabatically
from $t=-\infty$ so as not to disturb the system too far from
equilibrium in the process. If at $t=-\infty$ the system was in an
equilibrium state, then the condition of equilibrium
(\ref{equicond}) ensures that for $t<0$ there is a solution of the
equations of motion (\ref{eom1}) of the form
\begin{equation}
\delta_{\bk}(t) = \delta_{\bk}(0)\,e^{\epsilon t}\quad
\mathrm{for}\;\;t<0,\label{solless0}
\end{equation}
where $\delta_\bk(0)$ is related to $\eta_\bk$ through the
equations of motion for $t<0$. The advantage of the adiabatic
switching-on of the external source is that the time derivative of
the solution (\ref{solless0}) satisfies $\dot{\delta}_{\bk}(t<0)
\to 0$ as $\epsilon \to 0^+$.

Let us introduce auxiliary quantities $\Pi^R_{ab}(k,t-t')$ defined
as
\begin{equation}
 \Sigma^R_{ab}(k,t-t')=\partial_{t'}\Pi^R_{ab}(k,t-t'),
 \label{piret}
 \end{equation}
then upon using integration by parts, neglecting terms that vanish
in the adiabatic limit $\epsilon \to 0^+$, and taking
 $\Delta_0$ to be the equilibrium condensate and hence
$\mathcal{T}(\Delta_0,\Delta^\ast_0)=0$, we find the equations of
motion (\ref{eom1}) become
\begin{eqnarray}
&&[1+g\Pi^R_{11}(k,0)]\delta_\bk(t)+
g\Pi^R_{12}(k,0)\delta^\ast_{-\bk}(t)-g\int_0^t
dt'\big[\Pi^R_{11}(k,t-t')\dot{\delta}_\bk(t')+\Pi^R_{12}(k,t-t')
\dot{\delta}^\ast_{-\bk}(t')\big]=0,\nn\\
&&[1+g\Pi^R_{22}(k,0)]\delta^\ast_{-\bk}(t)+
g\Pi^R_{21}(k,0)\delta_\bk(t)-g\int_0^t
dt'\big[\Pi^R_{22}(k,t-t')\dot{\delta}^\ast_{-\bk}(t')+\Pi^R_{21}(k,t-t')
\dot{\delta}_\bk(t')\big]=0.\label{eom2l}
\end{eqnarray}
The above coupled equations of motion for the condensate
perturbation are now in the form of an \emph{initial value
problem} with initial conditions specified at $t=0$ and can be
solved by Laplace transform. Introducing a two-component
Nambu-Gor'kov spinor and its Laplace transform
\begin{eqnarray}\label{namgor}
&&\phi_\bk(t)= \begin{bmatrix}
\delta_\bk(t)\\
\delta^\ast_{-\bk}(t)
\end{bmatrix},\quad
\widetilde{\phi}_\bk(s)=\begin{bmatrix}
\widetilde{\delta}_\bk(s)\\
\widetilde{\delta}^\ast_{-\bk}(s)
\end{bmatrix},
\end{eqnarray}
where $s$ is the Laplace variable, one can rewrite the Laplace
transformed equations of motion in a compact matrix form as
\begin{equation}\label{laplaeqn}
\widetilde{G}^{-1}(k,s)\,\widetilde{\phi}_\bk(s)=\frac{1}{s}
\big[\widetilde{G}^{-1}(k,s)-\widetilde{G}^{-1}(k,0)\big]
\phi_\bk(0).
\end{equation}
In the above equation, $\widetilde{G}^{-1}(k,s)$ is the inverse
Green's function (matrix) to one-loop order expressed in terms of
the Laplace variable $s$
\begin{equation}
\widetilde{G}^{-1}(k,s)=\begin{bmatrix}
1+g\widetilde{\Sigma}_D(k,s) &
g(\Delta_0/\Delta_0^\ast)\widetilde{\Sigma}_O(k,s)\\
g(\Delta_0^\ast/\Delta_0)\widetilde{\Sigma}_O(k,s) &
1+g\widetilde{\Sigma}_D(k,-s)
\end{bmatrix},\label{eom3}
\end{equation}
where $\widetilde{\Sigma}_D$ and $\widetilde{\Sigma}_O$ are the
Laplace transforms of $\Sigma_D$ and $\Sigma_O$, respectively,
\begin{eqnarray}
&&\widetilde{\Sigma}_D(k,s)=\int^{+\infty}_{-\infty}
\frac{dk_0}{k_0-is}\,\left[\overline{A}(k,k_0)+\overline{S}(k,k_0)\right],\nn\\
&&\widetilde{\Sigma}_O(k,s)=\int^{+\infty}_{-\infty}
\frac{dk_0}{k_0-is}\widehat{A}(k,k_0). \label{sigmatilde}
\end{eqnarray}
The solution of (\ref{eom3}) reads
\begin{equation}
\widetilde{\phi}_\bk(s)=\frac{1}{s}\,[1-\widetilde{G}(k,s)\,
\widetilde{G}^{-1}(k,0)]\,\phi_\bk(0), \label{eom4}
\end{equation}
where
\begin{equation}
\widetilde{G}(k,s)=\frac{1}{\widetilde{D}(k,s)}\begin{bmatrix}
1+g\widetilde{\Sigma}_D(k,-s) &
-g(\Delta_0/\Delta_0^\ast)\widetilde{\Sigma}_O(k,s)\\
-g(\Delta_0^\ast/\Delta_0)\widetilde{\Sigma}_O(k,s) &
1+g\widetilde{\Sigma}_D(k,s)
\end{bmatrix},
\end{equation}
with the denominator $\widetilde{D}(k,s)$ given by
\begin{equation}
\widetilde{D}(k,s)=[1+g\widetilde{\Sigma}_D(k,s)]
[1+g\widetilde{\Sigma}_D(k,-s)]
-[g\widetilde{\Sigma}_O(k,s)]^2.\label{D}
\end{equation}

The real-time evolution of the condensate perturbation
$\phi_\bk(t)$ with an initial value $\phi_\bk(0)$ is now obtained
from the inverse Laplace transform
\begin{equation}\label{antilap}
\phi_\bk(t)= \int_\mathcal{B}\frac{ds}{2\pi i}\,e^{st}\,
\widetilde{\phi}_\bk(s),
\end{equation}
where the Bromwich contour $\mathcal{B}$ runs parallel to the
imaginary axis in the complex $s$ plane to the right of all the
singularities (poles and cuts) of $\widetilde{\phi}_\bk(s)$
\cite{ivp}. We note that there is \emph{no} isolated pole in
$\widetilde{\phi}_\bk(s)$ at $s=0$ since the residue vanishes.

\subsection{Self-energies in the long-wavelength, low-frequency limit}

In this article we are interested to study the relaxation of
long-wavelength, low-frequency fluctuations of the pair field,
hence our next task is to expand the self-energies as a function
of $k,\,s$ up to $\mathcal{O}(k^2,s^2)$. With $\xi_q=q^2/2m-\mu$
and $p=|\bq+ \bk|$ we use the following approximations
\begin{gather}
\xi_p \approx \xi_q + \delta \xi,\quad\delta \xi= v_\mathrm{F} k \cos\theta,\nn \\
E_p=E_q+ \xi_q \frac{\delta \xi}{E_q}+\frac{(\delta \xi)^2}{2}
\frac{|\Delta_0|^2}{E^3_q}, \label{aproximations}
\end{gather}
where $\theta$ is the angle between $\bhk$ and $\bhq$. We keep
only up to terms of $\mathcal{O}(k^2,s^2)$ in the Laplace
transform of the self-energies and obtain
\begin{eqnarray}
\widetilde{\Sigma}_D(k,s)&=&-\mathcal{N}(0)\int^{\omega_\mathrm{D}}_0
d\xi
\bigg\{\frac{1-2n_\mathrm{F}(E)}{2E}\bigg[\bigg(1+\frac{\xi^2}{E^2}\bigg)
\left(1-\frac{v_\mathrm{F}^2 k^2}{12E^2}-\frac{s^2}{4E^2}\right)-
\left(3-\frac{5\xi^2}{E^2}\right)\frac{v_\mathrm{F}^2 k^2\xi^2}{6E^4}\bigg]\nn\\
&&-\,\frac{\partial n_\mathrm{F}(E)}{\partial
E}\frac{|\Delta_0|^2}{E^2}\bigg[ \frac{v_\mathrm{F}^2
k^2\xi^2}{2E^4}+\left(1-\frac{3s^2}{2E^2}\right)\bigg(
1-\frac{isE}{2v_\mathrm{F}k\xi}\ln
\frac{is/v_\mathrm{F}k+\xi/E}{is/v_\mathrm{F}k-\xi/E}\bigg)\bigg]\bigg\}
+\cdots,\nn\\
\widetilde{\Sigma}_O(k,s)&=&\mathcal{N}(0)\int^{\omega_\mathrm{D}}_0
d\xi \frac{|\Delta_0|^2}{E^2}\bigg\{\frac{1-2n_\mathrm{F}(E)}{2E}
\bigg[1-\left(\frac{1}{4}-\frac{5\xi^2}{6E^2}\right)
\frac{v_\mathrm{F}^2k^2}{E^2}-\frac{s^2}{4E^2}\bigg]\nn\\
&&-\,\frac{\partial n_\mathrm{F}(E)}{\partial E}\bigg[
\frac{v_\mathrm{F}^2k^2}{2E^2}\left(\frac{1}{3}-\frac{\xi^2}{E^2}\right)-
\left(1-\frac{3s^2}{2E^2}+\frac{s^2}{\xi^2}\right)
\bigg(1-\frac{isE}{2v_\mathrm{F}k\xi}\ln
\frac{is/v_\mathrm{F}k+\xi/E}{is/v_\mathrm{F}k-\xi/E}\bigg)\bigg]\bigg\}+\cdots,
\label{sigmasexpands}
\end{eqnarray}
where $E=\sqrt{\xi^2+|\Delta_0|^2}$ and the dots stand for terms
of higher order in the ratios $k/|\Delta_0|, s/|\Delta_0|$. We
note that to this order both $\widetilde{\Sigma}_D(k,s)$ and
$\widetilde{\Sigma}_O(k,s)$ are even functions of the Laplace
variable $s$. This important feature leads to the decoupling of
the phase and amplitude fluctuations of the pair field in the
equations of motion, as will be shown below.

\subsection{One-loop effective action}

The full equations of motion in real time (\ref{eom1}) and their
Laplace transform around the equilibrium state (\ref{laplaeqn})
allow us to obtain at once the \emph{retarded effective action} in
Fourier space. The retarded Green's function kernel is obtained
from $\widetilde{G}^{-1}(k,s)$ in (\ref{laplaeqn}) by the analytic
continuation in the complex $s$ plane $s\to -i\omega+0^+$. In
terms of the Fourier transform (retarded) of the two-component
Nambu-Gor'kov spinor and the tadpole spinor
\begin{equation}
\mathcal{T}_\bk(\omega) =
\begin{bmatrix}
\mathcal{T}(\Delta_0,\Delta^\ast_0) \\
\mathcal{T}^\ast(\Delta_0,\Delta^\ast_0)
\end{bmatrix}
\delta^{(3)}(\bk)\delta(\omega),\label{tadpolespinor}
\end{equation}
the retarded one-loop effective action to quadratic order in the
fluctuations is therefore given by
\begin{equation}
S[\delta,\delta^\ast]= S[0,0]+\frac{1}{2g}\int d^3k d\omega
\left[\phi^{\dagger}_{-\bk}(-\omega)G^{-1}(k,\omega)\phi_\bk(\omega)+
2\phi^{\dagger}_{-\bk}(-\omega)\mathcal{T}_\bk(\omega)\right],
\label{effaction}
\end{equation}
where $G^{-1}(k,\omega)=\widetilde{G}^{-1}(k,s=-i\omega+0^+)$ and
$S[0,0]$ is a function of $\Delta_0$, $\Delta^\ast_0$ such that
\begin{equation}
\frac{\partial S[0,0]}{\partial \Delta^\ast_0} =
\mathcal{T}(\Delta_0,\Delta^\ast_0), \quad \frac{\partial
S[0,0]}{\partial \Delta_0}=
\mathcal{T}^\ast(\Delta_0,\Delta^\ast_0).\label{effderi}
\end{equation}
Obviously, variational derivatives with respect to
$\delta,\,\delta^\ast$ reproduce the retarded equations of motion.
We identify (\ref{effaction}) with the one-loop effective action
quadratic in the fluctuations.

\section{Ward identity and static self-energies}\label{sec:WI}

The Ward identities, a consequence of the underlying (global)
gauge invariance, are an integral part of the program to establish
a connection with the Ginzburg-Landau description. Furthermore the
equations of motion \emph{must} fulfill these for consistency. In
this section we show how the  Ward identities emerge directly from
the method described above used to obtain the equations of motion.

 A straightforward diagrammatic analysis with the
Feynman rules described above reveals that the generic structure
of the equations of motion obtained via the tadpole method remains
the same to all orders in perturbation theory. When combined with
the transformation properties of $\Delta_0$, $\delta$ and $\chi$
under the gauge transformation (\ref{gauge}), this general form of
the equations of motion allows to derive to all orders in
perturbation theory the Ward identity for the tadpole
$\mathcal{T}(\Delta_0,\Delta^\ast_0)$.

First, consider the case in which $\delta,\,\eta=0$ the tadpole
condition $\langle\chi^+(0)\rangle_\eta=0$ leads to
$\mathcal{T}(\Delta_0,\Delta^\ast_0) =0$, which is the equilibrium
condition for the homogeneous condensate. For a space-time
independent shift of the condensate  $\Delta_0 \to
\Delta_0+\delta$ induced by a space-time independent source
$\eta$, the tadpole condition now leads to
$\mathcal{T}(\Delta_0+\delta,\Delta_0^\ast+\delta^\ast) =-\eta$,
which upon expanding to linear order in $\delta$ and $\delta^\ast$
becomes
\begin{equation}
\mathcal{T}(\Delta_0,\Delta^\ast_0)+\frac{\partial
\mathcal{T}(\Delta_0,\Delta^\ast_0)}{\partial
\Delta_0}\,\delta+\frac{\partial
\mathcal{T}(\Delta_0,\Delta^\ast_0)}{\partial
\Delta_0^\ast}\,\delta^\ast=-\eta.\label{spindeom}
\end{equation}
We now compare (\ref{spindeom}) with the first equation of motion
given in (\ref{eom1}), which has the same generic structure as the
full equation of motion obtained to all orders in perturbation
theory, for space-time independent $\delta$ and $\delta^\ast$.
Comparing the coefficients of $\delta,\delta^\ast$ we recognize
immediately that
\begin{eqnarray}
\frac{\partial \mathcal{T}(\Delta_0,\Delta^\ast_0)}{\partial
\Delta_0}
&=&\frac{1}{g}+\Sigma_{11}(k=0,\omega=0), \nn\\
\frac{\partial \mathcal{T}(\Delta_0,\Delta^\ast_0)}{\partial
\Delta_0^\ast} &=& \Sigma_{12}(k=0,\omega=0), \label{selftads}
\end{eqnarray}
which relate the self-energies at \emph{zero frequency} and
\emph{zero momentum} to the derivatives of the tadpole with
respect to the condensate and is valid to \emph{all orders in
perturbation theory}.

The second important ingredient and which stems from the equations
of motion (\ref{eom1}) is that under a global gauge (phase)
transformation (\ref{gauge}) the tadpole
$\mathcal{T}(\Delta_0,\Delta^\ast_0)$ transforms just as $\delta$,
$\Delta_0$ and $\eta$, i.e.,
$\mathcal{T}(e^{i\theta}\Delta_0,e^{-i\theta}\Delta_0^\ast)
=e^{i\theta}\mathcal{T}(\Delta_0,\Delta^\ast_0)$. Taking the gauge
parameter $\theta$ to be infinitesimal and comparing the linear
terms in $\theta$, we find to \emph{all orders in perturbation
theory} the Ward identity for the tadpole
\begin{equation}
\mathcal{T}(\Delta_0,\Delta^\ast_0) =\frac{\partial
\mathcal{T}(\Delta_0,\Delta^\ast_0)}{\partial \Delta_0} ~\Delta_0
-\frac{\partial \mathcal{T}(\Delta_0,\Delta^\ast_0)}{\partial
\Delta_0^\ast} \Delta_0^\ast.\label{Ward}
\end{equation}
Therefore upon combining (\ref{selftads}) and the Ward identity
(\ref{Ward}), we obtain  an alternative statement of the   Ward
identity which is an \emph{exact} relationship between the tadpole
and the self-energies at zero frequency and momentum
\begin{equation}
\Delta_0 \left[\frac{1}{g}+\Sigma_{11}(0,0)\right]
-\Delta_0^\ast\,\Sigma_{12}(0,0)=
\mathcal{T}(\Delta_0,\Delta^\ast_0),\label{exactrel}
\end{equation}
where $\Sigma_{ab}(0,0)$ stands for $\Sigma_{ab}(k=0,\omega=0)$
for notational simplicity. Above the critical temperature,
\emph{both} the tadpole and the condensate vanish thus the above
equation becomes a trivial identity. However, below the critical
temperature $\Delta_0 \neq 0$ and hence (\ref{exactrel}) leads to
\begin{equation}
\frac{g\mathcal{T}(\Delta_0,\Delta^\ast_0)}{\Delta_0}=1+g\left[
\Sigma_{11}(0,0)-\frac{\Delta_0^\ast}{\Delta_0}
\Sigma_{12}(0,0)\right].\label{WI}
\end{equation}

It is customary to choose the condensate to be real by redefining
its phase via the gauge transformation (\ref{gauge}). However, as
argued above, for a condensate with an arbitrary phase, the
anomalous self-energy $\Sigma_{12}$ \emph{must be} proportional to
$\Delta_0^2$ since in the equation of motion it multiplies
$\delta^\ast$, which transforms under gauge transformations just
as $\Delta_0^\ast$. This fact can be seen explicitly at the
one-loop order in the expressions for the respective anomalous
self-energy in (\ref{sigmaDO}) as well as in (\ref{sigmas}) with
the spectral functions given by (\ref{spectral}). Since the
product $\Sigma_{12}\delta^\ast$ must transform just as $\delta$
or $\Delta_0$, therefore the phase $\Delta_0^\ast/\Delta_0$
cancels the phase of $\Delta^2_0$ in $\Sigma_{12}$. In terms of
the \emph{gauge invariant} self-energies $\Sigma_D$ and
$\Sigma_O$, the Ward identity (\ref{WI}) can be cast into an
explicit gauge invariant form
\begin{equation}
\frac{g\mathcal{T}(\Delta_0,\Delta^\ast_0)}{\Delta_0}=1+
g\Sigma_D(0,0)-g\Sigma_O(0,0).\label{WI2}
\end{equation}
We emphasize that the Ward identity (\ref{WI}) or (\ref{WI2}) is
an exact relationship valid \emph{in} or \emph{out} of equilibrium
(corresponding to when the tadpole
$\mathcal{T}(\Delta_0,\Delta^\ast_0)$ vanishes or not,
respectively). In equilibrium the equilibrium condition
$\mathcal{T}(\Delta_0,\Delta^\ast_0)=0$ implies that the
self-energies in the static limit must satisfy the relation
\begin{equation}
1+g\Sigma_D(0,0)-g\Sigma_O(0,0)=0. \label{hp}
\end{equation}

The long-wavelength limit of the \emph{static} self-energies is
\emph{defined} as
\begin{eqnarray}
&&\Sigma_D(0,0)=\lim_{k\to 0}\lim_{s\to 0} \Sigma_D(k,s), \nn\\
&&\Sigma_O(0,0)=\lim_{k\to 0}\lim_{s\to
0}\Sigma_O(k,s),\label{statlimit}
\end{eqnarray}
which, using (\ref{sigmasexpands}), are found to one-loop order
given by
\begin{eqnarray}
&&\Sigma_D(0,0)= -\mathcal{N}(0)
\int_0^{\omega_\mathrm{D}}d\xi\,\bigg[\frac{1-2n_\mathrm{F}(E)}{2E}
\left(1+\frac{\xi^2}{E^2}\right)-\frac{\partial
n_\mathrm{F}(E)}{\partial E}
\frac{|\Delta_0|^2}{E^2}\bigg],\nn\\
&&\Sigma_O(0,0)= \mathcal{N}(0)\int_0^{\omega_\mathrm{D}}d\xi\,
\frac{|\Delta_0|^2}{E^2}\bigg[\frac{1-2n_\mathrm{F}(E)}{2E}
+\frac{\partial n_\mathrm{F}(E)}{\partial E}\bigg].\label{slimit}
\end{eqnarray}
The opposite limit $\lim_{s\to 0} \lim_{k\to 0}\Sigma_{D,O}(k,s)$
yields the first terms in the expressions (\ref{slimit}) but
\emph{not} the last terms proportional to $\partial
n_\mathrm{F}(E)/\partial E$. This is a consequence of the
non-analyticity of the Landau damping contributions to the
self-energies.

Using the expression for the one-loop tadpole given by
(\ref{tadpoleterm2}) it becomes clear that the identities
(\ref{selftads}) are fulfilled by the precise order of limits
determined by \eqref{slimit}, i.e., the long-wavelength limit of
the \emph{static} ($s=0$) self-energies.

We now show explicitly that the Ward identity (\ref{WI2}) is
fulfilled to one-loop order. The tadpole to one-loop order is
given by (\ref{tadpoleterm2}). From the explicit form of the
one-loop self-energies in the static limit (\ref{slimit}),  find
that
\begin{equation}
\Sigma_D(0,0)-\Sigma_O(0,0)=
-\mathcal{N}(0)\int_0^{\omega_\mathrm{D}}
\frac{d\xi}{E}\,[1-2n_\mathrm{F}(E)],\label{NLrels}
\end{equation}
which is an \emph{exact} relationship to one-loop order, obtained
for arbitrary $\Delta_0$.

Upon collecting the above results, we find to one-loop order that
\begin{eqnarray}
\frac{g\mathcal{T}(\Delta_0,\Delta^\ast_0)}{\Delta_0}&=&
1+g\Sigma_D(0,0)-g\Sigma_O(0,0)\nn\\
&=&1-g\mathcal{N}(0)\int_0^{\omega_\mathrm{D}}
\frac{d\xi}{E}\,[1-2n_\mathrm{F}(E)].\label{WI1lup}
\end{eqnarray}
Therefore, the Ward identity (\ref{WI2}) is manifestly fulfilled
to one-loop order. This is an important advantage of the tadpole
method of nonequilibrium field theory: the equations of motion
obtained at a given order in the loop expansion are causal and
guaranteed to fulfill the corresponding Ward identities to that
order.

We highlight that whereas the Ward identity is independent of the
limits, $k,s \to 0$, the individual self-energies have different
limits because of the non-analyticity of the Landau damping
contribution. The relationship between the self-energies and the
tadpole \eqref{selftads} is only valid in the static limit
\eqref{slimit}.

The results of this section, namely the Ward identity which
relates the self-energies to derivatives of the tadpole will play
an important role in the derivation of the effective
Ginzburg-Landau theory.

\section{Time-dependent Ginzburg-Landau Theory}\label{sec:TDGL}

We are now in position to establish contact with the
Ginzburg-Landau description of the long-wavelength, low-frequency
excitations near the critical point. The Ginzburg-Landau
description in terms of a functional of the order parameter is
valid near the critical region when $|T-T_c| \ll T_c,
\Delta(T)\simeq \left[T_c(T_c-T)\right]^{1/2}\ll T_c$. The link
between the Ginzburg-Landau (GL) theory and the microscopic theory
of superconductivity in the static limit was established by
Gor'kov \cite{gorkov59}.

Identifying the expectation value of the pair field
$\Delta\equiv\langle\Delta(\xt)\rangle$ as the complex order
parameter, the free energy in Ginzburg-Landau theory for an
homogeneous order parameter (absence of gradients) is given by
\begin{equation}
\mathcal{F}(\Delta,\Delta^\ast) = a(T)
|\Delta|^2+\frac{b}{2}|\Delta|^4,\label{freeenergyLG}
\end{equation}
where near the critical temperature $a(T)\propto (T-T_c)$.

The linearized equation of motion of the small amplitude
fluctuation of the condensate can be obtained by expanding the
free energy around the equilibrium value of gap function; namely
$\Delta_0$ and $\Delta_0^\ast$. Writing $\Delta = \Delta_0 +
\delta$ and $\Delta^\ast=\Delta_0^\ast + \delta^\ast$, keeping
only quadratic terms in the fluctuations:
\begin{eqnarray}
&&\mathcal{F}(\Delta,\Delta^\ast)=
\mathcal{F}(\Delta_0,\Delta^\ast_0) + \delta\frac{\partial
\mathcal{F}}{\partial \Delta} + \delta^\ast \frac{\partial
\mathcal{F}}{\partial \Delta^\ast} + \delta \delta^\ast
\frac{\partial^2 \mathcal{F}}{\partial \Delta\partial
\Delta^\ast}+\frac{\delta^2}{2}\frac{\partial^2
\mathcal{F}}{\partial \Delta^2} + \frac{\delta^{\ast
2}}{2}\frac{\partial^2 \mathcal{F}}{\partial\Delta^{\ast 2}}.
\end{eqnarray}

The linearized equation of motion for both $\delta$ and
$\delta^\ast$ can then be obtained by minimizing $\mathcal{F}$
with respect to $\delta^\ast$ and $\delta$ respectively and are
given by
\begin{gather}
\Sigma^\mathrm{GL}_{11} \,\delta+\Sigma^\mathrm{GL}_{12}
\delta^\ast +{\mathcal
T}^\mathrm{GL}(\Delta_0,\Delta^\ast_0)= 0,\nn \\
\Sigma^\mathrm{GL}_{22}\delta^\ast+\Sigma^\mathrm{GL}_{21}\delta+[{\mathcal
T}^{\mathrm{GL}}(\Delta_0,\Delta^\ast_0)]^\ast=0,\label{eqnofmotLG}
\end{gather}
with
\begin{gather}
{\mathcal
T}^\mathrm{GL}(\Delta_0,\Delta^\ast_0)=\Delta_0\left[a(T)+
b|\Delta_0|^2\right], \nn\\
\Sigma^\mathrm{GL}_{11}= [\Sigma^\mathrm{GL}_{22}]^\ast=a(T)+2b
|\Delta_0|^2
\equiv \frac{\partial{\mathcal T}^\mathrm{GL}}{\partial\Delta_0}, \nn \\
\Sigma^\mathrm{GL}_{12}=[\Sigma^\mathrm{GL}_{21}]^\ast=b
\Delta^2_0 \equiv \frac{\partial{\mathcal
T}^\mathrm{GL}}{\partial\Delta^\ast_0}.\label{identitiesLG}
\end{gather}

The gap equation obtained from the Ginzburg-Landau free energy
(\ref{freeenergyLG}) is
\begin{equation}
{\mathcal
T}^\mathrm{GL}(\Delta_0,\Delta^\ast_0)=0,\label{gapeqnLG}
\end{equation}
which determines the equilibrium value of the order parameter for
$\delta=\delta^\ast=0$. When the gap equation (\ref{gapeqnLG}) is
fulfilled, the matrix of the second derivatives of the free energy
(\ref{freeenergyLG}) with respect to $\delta,\delta^\ast$ has a
zero eigenvalue with eigenvector determined by the relation
\begin{equation}
\delta^\ast = - \frac{\Delta^\ast_0}{\Delta_0}\delta.
\end{equation}
Obviously this eigenvector with zero eigenvalue corresponds to a
phase fluctuation and is the Goldstone boson, or the
Anderson-Bogoliubov-Goldstone mode, associated with the broken
global gauge symmetry.

The similarity between the equations of motion (\ref{eqnofmotLG})
and (\ref{eom1}) for an homogeneous perturbation with vanishing
external sources, as well as the similarities between the
identities (\ref{identitiesLG}) and (\ref{selftads}),
(\ref{exactrel}) are now obvious and suggest the following
identification
\begin{eqnarray}
\mathcal{T}(\Delta_0,\Delta^\ast_0) & \Longleftrightarrow &
\mathcal{T}^\mathrm{GL}(\Delta_0,\Delta^\ast_0),\nn \\
\frac{1}{g}+\Sigma_{11}(0,0)  & \Longleftrightarrow &
\Sigma^\mathrm{GL}_{11}, \nn \\
\Sigma_{12}(0,0) & \Longleftrightarrow &
\Sigma^\mathrm{GL}_{12},\label{identi}
\end{eqnarray}
where the self-energies on the left-hand side of the relations are
understood as the long-wavelength limit of the \emph{static} self
energies as discussed in Sec.~\ref{sec:WI}. This similarity can be
put on firmer footing by expanding the one-loop expression for the
tadpole $\mathcal{T}(\Delta_0,\Delta^\ast_0)$ (\ref{tadpoleterm2})
in terms of a power series expansion in $\Delta_0,\Delta^\ast_0$
and keeping up to third order consistently with the
Ginzburg-Landau expansion.

Using the form of the tadpole given in (\ref{tadpoleterm2}) and
the results obtained in Ref.~\onlinecite{ketterson}, we find
($\hbar=k_\mathrm{B}=1$)
\begin{equation}\label{LGexp}
g\mathcal{T}(\Delta_0,\Delta^\ast_0)= \left[1-g\mathcal{N}(0)
\ln\left(\frac{1.14  \omega_\mathrm{D}}{T}\right)\right]\Delta_0+
\frac{7g\mathcal{N}(0)\zeta(3)}{8 \pi^2 T^2_c}\Delta_0
|\Delta_0|^2 + \mathcal{O}(\Delta_0^5),
\end{equation}
where $\zeta$ is the Riemann zeta function with $\zeta(3)=1.202$
and $T_c= 1.14\,\omega_\mathrm{D}\,e^{-1/g\mathcal{N}(0)}$. For
$T\approx T_c$ the coefficient of the linear term can be expanded
as $g\mathcal{N}(0)(T-T_c)/T_c$, therefore from the relations
(\ref{selftads}) and the definitions (\ref{sigmaDO}) we obtain
\begin{eqnarray}\label{LGcoefs}
1+g\Sigma_{D}(0,0) & =
&g\mathcal{N}(0)\left(\frac{T-T_c}{T_c}\right)+\frac{7 g
\mathcal{N}(0)\zeta(3)}{4\pi^2  T^2_c} |\Delta_0|^2, \nn \\
g\Sigma_{O}(0,0) & = & \frac{7 g \mathcal{N}(0)\zeta(3)}{8 \pi^2
T^2_c} |\Delta_0|^2.
\end{eqnarray}

We now focus on the long-wavelength, low-frequency self-energies
in the Ginzburg-Landau regime characterized by $v_\mathrm{F}k,
s\ll |\Delta_0| \ll T$. This regime corresponds to the description
of the long-wavelength, low-frequency excitations near the
critical point and will allow a consistent analysis of the
real-time dynamics of (small) phase and amplitude fluctuations. We
begin by writing the long-wavelength, low-frequency self-energies
as [see (\ref{sigmasexpands})]
\begin{eqnarray}
\widetilde{\Sigma}_D(k,s) &=& \Sigma_D(0,0)+ I(k,s),\nn\\
\widetilde{\Sigma}_O(k,s) &=& \Sigma_O(0,0)+ J(k,s),
\label{sigDOks}
\end{eqnarray}
with $\Sigma_{D,O}(0,0)$ given by (\ref{slimit}). Since the
self-energies are dimensionless functions of their arguments, the
expansion in terms of $k$  and $s$ must involve ratios of these
variables and the typical scales in the integrals. There are two,
widely separated scales in the Ginzburg-Landau regime, namely,
$T\approx T_c$ and $|\Delta_0|$ with $|\Delta_0|/T \ll 1$. The
expressions for $I,J$ feature energy denominators that would lead
to infrared divergences if $|\Delta_0|$ is set to zero. These
divergences reflect the presence of inverse powers of $|\Delta_0|$
in the expansions.

The leading terms in the expansion can be extracted by rescaling
\begin{equation}
\xi=|\Delta_0| z,\quad E=|\Delta_0| \epsilon,\label{rescale}
\end{equation}
with $\epsilon= \sqrt{z^2+1}$ and introducing the dimensionless
variables
\begin{equation}
x=\frac{|\Delta_0|}{T},\quad\kappa=\frac{v_\mathrm{F}k}{|\Delta_0|},\quad
\sbar=\frac{s}{v_\mathrm{F} k},
\end{equation}
in the integrals for $I(k,s)$ and $J(k,s)$. We obtain
\begin{equation}
I(k,s) = \mathcal{N}(0)\,x\left[\kappa^2 I_a(x)+ \kappa^2 \sbar^2
I_b(\sbar,x)+i\sbar I_c(\sbar,x)\right],\label{I}
\end{equation}
where
\begin{eqnarray}
I_a(x) & = &\frac{1}{4}\int^{\infty}_0 dz
\left\{\frac{2T(x,z)}{3x\epsilon^3}
\left[\frac{2z^2}{\epsilon^4}-\frac{z^2-1}{4\epsilon^2}
\left(1+\frac{z^2}{\epsilon^2}\right)\right]
-\frac{C(x,z)z^2}{2\epsilon^6})\right\},  \nn \\
I_b\left(\sbar,x\right)& = &\frac{1}{4}\int^{\infty}_0
dz\left\{\frac{T(x,z)}{2x\epsilon^3}
\left(1+\frac{z^2}{\epsilon^2}\right)+
\frac{3C(x,z)}{2\epsilon^4}\left[1-\frac{i\sbar\epsilon }{2z}
\ln\frac{i\sbar+z/\epsilon }{i\sbar-z/\epsilon }
\right]\right\},\nn \\
I_c \left(\sbar,x\right) & = &\frac{1}{4}\int^{\infty}_0
dz\left[\frac{C(x,z)}{2z\epsilon } \ln\frac{i\sbar+z/\epsilon
}{i\sbar-z/\epsilon }\right],\label{Iintegrals}
\end{eqnarray}
with
\begin{equation}
T(x,z)=\tanh\left(\frac{x\epsilon }{2}\right),\quad
C(x,z)=\cosh^{-2}\left(\frac{x\epsilon }{2}\right).
\end{equation}
Similarly, for $J(k,s)$ we obtain
\begin{equation}
J(k,s) = \mathcal{N}(0)\,x\left[\kappa^2 J_a(x)+ \kappa^2 \sbar^2
J_b(\sbar,x)+i\sbar J_c(\sbar,x)\right], \label{J}
\end{equation}
where
\begin{eqnarray}
J_a(x) & = & \frac{1}{24}\int^{\infty}_0 dz \left[T(x,z)\frac{7z^2
-3}{x\epsilon^7}-C(x,z)\frac{2z^2-1}{\epsilon^6}\right], \nn \\
J_b \left(\sbar,x \right)& =&-\frac{1}{8}\int^{\infty}_0 dz
\left\{\frac{T(x,z)}{x\epsilon^5}-
C(x,z)\frac{2z^2-1}{z^2\epsilon^4} \left[1-\frac{i\sbar\epsilon
}{2z} \ln\frac{i\sbar+z/\epsilon }{i\sbar-z/\epsilon }
\right] \right\}, \nn \\
J_c\left(\sbar,x\right)& = &
I_c\left(\sbar,x\right).\label{Jintegrals}
\end{eqnarray}

An important consequence of the long-wavelength, low-frequency
expansion of the self-energies in (\ref{sigDOks}) together with
(\ref{Iintegrals}), (\ref{Jintegrals}) is that to lowest order the
self-energies are \emph{even} functions of the Laplace variable
$s$. This important aspect leads to the decoupling of the phase
and amplitude fluctuations in the equations of motion, as can be
seen as follows.

If we write the fluctuation of the order parameter around the
space-time constant equilibrium solution of the tadpole equation
(\ref{equicond}) in the form
\begin{eqnarray}
\Delta(x,t)&=&\Delta_0+\delta(x,t)\nn\\
&\equiv&[\rho_0+\delta \rho(x,t)]\,e^{i\theta_0+i\delta
\theta(x,t)},
\end{eqnarray}
where $\delta\rho(x,t)$ and $\delta \theta(x,t)$, respectively,
are identified as the amplitude and phase fluctuations of the
order parameter $\Delta(x,t)$. It is convenient to introduce the
following projection vectors
\begin{eqnarray}
\bp& = & \frac{1}{\sqrt{2}i}
\begin{bmatrix}
1 \\-\frac{\Delta^\ast_0}{\Delta_0}
\end{bmatrix},\quad
\bp^\dagger=-\frac{1}{\sqrt{2}i}
\left[1,-\frac{\Delta_0}{\Delta^\ast_0}\right],\nn \\
\ba& = & \frac{1}{\sqrt{2}}
\begin{bmatrix}
1 \\ \frac{\Delta^\ast_0}{\Delta_0}
\end{bmatrix},\quad
\ba^\dagger=\frac{1}{\sqrt{2}}\left[1,\frac{\Delta_0}{\Delta^\ast_0}\right],
\label{projectors}
\end{eqnarray}
in terms of these vectors and the Nambu-Gor'kov spinors
(\ref{namgor}), the phase and amplitude fluctuations can be
written as
\begin{equation}
\delta\theta(x,t) = \frac{1}{\sqrt{2}\Delta_0} \bp^\dagger\cdot
\phi(x,t),\quad \frac{\delta \rho(x,t)}{\rho_0}=
\frac{1}{\sqrt{2}\Delta_0} \ba^\dagger\cdot \phi(x,t),
\label{phaseamp}
\end{equation}
or, equivalently,
\begin{equation}
\widetilde{\phi}_\bk(s)=\sqrt{2}\Delta_0\left[\widetilde{\delta
\theta}_\bk(s) \bp+ \frac{\widetilde{\delta
\rho}_\bk(s)}{\rho_0}\ba\right].
\end{equation}

In terms of the Laplace transforms
$\widetilde{\delta\theta}_\bk(s)$ and
$\widetilde{\delta\rho}_\bk(s)$, the equations of motion
(\ref{laplaeqn}) now become
\begin{eqnarray}\label{paeqn}
\left[1+g\widetilde{\Sigma}_D(k,s)-g\widetilde{\Sigma}_O(k,s)
\right]\widetilde{\delta \theta}_\bk(s) = \frac{g}{s}\left[\left(
\widetilde{\Sigma}_D(k,s)-\widetilde{\Sigma}_O(k,s)\right)-\left(
\widetilde{\Sigma}_D(k,0)-\widetilde{\Sigma}_O(k,0)\right) \right]
\delta \theta_\bk(0), \nn \\
\left[1+g\widetilde{\Sigma}_D(k,s)+g\widetilde{\Sigma}_O(k,s)
\right]\widetilde{\delta \rho}_\bk(s) = \frac{g}{s}\left[\left(
\widetilde{\Sigma}_D(k,s)+\widetilde{\Sigma}_O(k,s)\right)-\left(
\widetilde{\Sigma}_D(k,0)+\widetilde{\Sigma}_O(k,0)\right) \right]
\delta \rho_\bk(0).
\end{eqnarray}
The decoupling of the equations of motion between phase and
amplitude fluctuations is a direct consequence of the fact that
self-energies are even functions of the Laplace variable to this,
lowest order. In particular the equality
$\Sigma_D(k,s)=\Sigma_D(k,-s)$ to this order guarantees that the
vectors $\bp$ and $\ba$ are eigenvectors of the matrix
$\widetilde{G}^{-1}(k,s)$ in (\ref{laplaeqn}). The right-hand
sides of (\ref{paeqn}) are identified as the Laplace transform of
the source terms $J_{\theta}$, $J_{\rho}$, respectively, that
serve the purpose of setting up the initial value problem.

In particular to lowest order in the long-wavelength expansion,
and after the analytic continuation $s\to -i\omega+0^+$, the
\emph{retarded} one-loop effective action for fluctuations around
the equilibrium solution is given, up to a constant, by
\begin{eqnarray}
S_\mathrm{eff}\left[\delta \theta, \delta \rho\right] &=&
\frac{|\Delta_0|^2}{2g} \int d^3k
d\omega\left\{\delta\theta_{-\bk}(-\omega)\left[1+g\Sigma_D(k,\omega)
-g\Sigma_O(k,\omega)\right]\delta\theta_\bk(\omega)+2
J_{\theta,-\bk}(-\omega)\delta\theta_\bk(\omega) \right\}\nn \\
&&+\,\frac{1}{2g} \int d^3k
d\omega\left\{\delta\rho_{-\bk}(-\omega)\left[1+g
\Sigma_D(k,\omega)+g\Sigma_O(k,\omega)\right]\delta
\rho_\bk(\omega)+2
J_{\rho,-\bk}(-\omega)\delta\rho_\bk(\omega)\right\},\label{effacphaseamp}
\end{eqnarray}
where
$\Sigma_{D,O}(k,\omega)=\widetilde{\Sigma}_{D,O}(k,s=-i\omega+0^+)$
and the momentum integral must be understood to be restricted near
the Fermi surface. Expanding the self-energies as in
(\ref{sigDOks}) together with (\ref{Iintegrals}),
(\ref{Jintegrals}) and using the Ward identities (\ref{WI2}) valid
in equilibrium, one can easily show that the phase fluctuations
are Goldstone modes. Using the explicit expressions for $I(k,s)$
and $J(k,s)$, we find that the effective action for the phase
fluctuation is identical to that obtained in
Refs.~\onlinecite{aitchison97,aitchison95} to quadratic order,
while the effective action for the amplitude fluctuation is a new
contribution.

The approach followed here, based directly on the equations of
motion in real time also allows us to obtain the effective action
for the amplitude fluctuations on the same footing.

%\subsection{Spectral densities and real-time evolution}
\subsection{Phase fluctuations}

The initial value problem for the phase and amplitude fluctuations
described by the equations of motions (\ref{paeqn}) can now be
studied straightforwardly. For the phase fluctuation the inverse
Laplace transform is obtained by integrating $\widetilde{\delta
\theta}_\bk(s)$ along the Bromwich contour in the complex $s$
plane parallel to the imaginary axis and to the right of all the
singularities of the Laplace transform
\begin{equation}
\frac{\widetilde{\delta\theta}_\bk(s)}{{\delta\theta}_\bk(0)}=\frac{1}{s}
\frac{F(\sbar,x)}{F(\sbar,x)+f(x)},\label{laplatheta}
\end{equation}
where
\begin{eqnarray}
&&f(x)=I_a(x)-J_a(x),\nn\\
&&F(\sbar,x)=\sbar^2\left[I_b\left(\sbar,x\right)-
J_b\left(\sbar,x\right)\right].
\end{eqnarray}

\begin{figure}[t]
\includegraphics[width=3.5in,keepaspectratio=true]{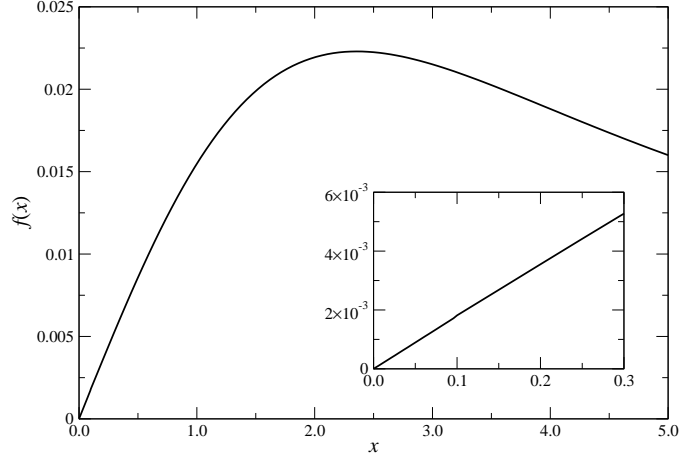}
\caption{The full $x$ dependence of $f(x)$ is plotted for $0<x<5$.
The inset shows $f(x)$ in the Ginzburg-Landau regime $x\ll 1$.}
\label{fig:IaJa}
\end{figure}

In the Ginzburg-Landau regime $x\ll 1$, we find from
(\ref{Iintegrals}) and (\ref{Jintegrals}) that
\begin{eqnarray}
f(x) &\simeq& -\frac{x}{12}\int^\infty_0
\frac{dz}{z^3}\left[\tanh\frac{z}{2}-\frac{z}{2}\right]+\mathcal{O}(x^2)\nn\\
&=& \frac{7\zeta(3)}{48\pi^2}x+\mathcal{O}(x^2). \label{IaJa}
\end{eqnarray}
Figure~\ref{fig:IaJa} displays a numerical evaluation of $f(x)$
with the full $x$ dependence. It shows clearly that $f(x)\ll 1$
and that $f(x)$ is accurately described by (\ref{IaJa}) in the
Ginzburg-Landau regime.

To leading order in $x$, we can set $x\to 0$ in the arguments when
evaluating $F(\sbar,x)$. After the analytic continuation $s\to
-i\omega+0^+$ and introducing the dimensionless variable
\begin{equation}
\alpha= \frac{\omega}{v_\mathrm{F} k},
\end{equation}
we find in the Ginzburg-Landau regime that the real and imaginary
parts of $F(\sbar=-i\alpha+0^+,0)$ are, respectively, given by
(here we have set $x\to 0$)
\begin{eqnarray}
F_\mathrm{R}(\alpha) & = & \frac{\alpha^2}{8} \int^{\infty}_0
\frac{dz}{\epsilon^2} \left[1+ \frac{1}{z^2} \left(1-\frac{\alpha
\epsilon }{2z}\ln\left|\frac{\alpha+
\frac{z}{\epsilon }}{\alpha-\frac{z}{\epsilon }}\right| \right)\right]\nn \\
&\stackrel{\alpha\ll 1}\simeq&\frac{\pi^2|\alpha|^3}{64},\nn\\
F_\mathrm{I}(\alpha) &=& \frac{\pi\alpha^3}{16}\Theta(1-\alpha^2)
\int^{\infty}_{z_\mathrm{min}}\frac{dz}{z^3\epsilon }\nn\\
&\stackrel{\alpha\ll 1}\simeq&\frac{\pi\alpha}{32},
\label{realimag}
\end{eqnarray}
where $z_\mathrm{min}=\sqrt{\alpha^2/(1-\alpha^2)}$ and we have
set $x\to 0$ in the argument of $F(\sbar,x)$. These expressions
agree with those found by Aitchison et al.\ \cite{aitchison97}
in the limit $x\to 0$ \cite{comparison1}.

\begin{figure}[t]
\includegraphics[width=3.5in,keepaspectratio=true]{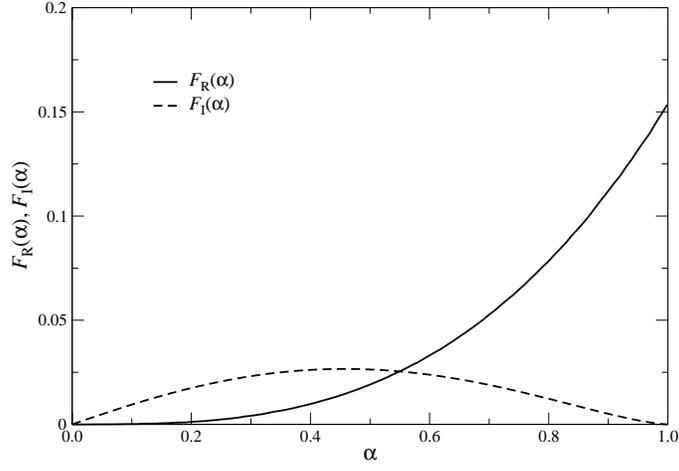}
\caption{Real and imaginary parts of $F(\alpha)$.}
\label{fig:realimagparts}
\end{figure}

The real [$F_\mathrm{R}(\alpha)$] and imaginary
[$F_\mathrm{I}(\alpha)$] parts of $F(\alpha)$ for $0\le\alpha\le
1$ are displayed in Fig.~\ref{fig:realimagparts}. The real part is
a monotonically increasing function of $\alpha$ while the
imaginary part only has contribution in the Landau damping cut
$-1<\alpha<1$ corresponding to
$-v_\mathrm{F}k<\omega<v_\mathrm{F}k$.

Isolated \emph{real} poles of the Laplace transform, describing
stable quasiparticle excitations, correspond to the solutions of
the following equations
\begin{equation}
F_\mathrm{R}(\alpha)-f(x)=0,\quad F_\mathrm{I}(\alpha)= 0.
\label{realpole}
\end{equation}
It is clear from (\ref{IaJa}) and Fig.~\ref{fig:realimagparts}
that the first equation in (\ref{realpole}) can be fulfilled for
$f(x)\ll 1$ only in a region where $F_\mathrm{I}(\alpha)\neq 0$.
Therefore in the Ginzburg-Landau regime the long-wavelength,
low-frequency Laplace transform (\ref{laplatheta}) has \emph{no}
isolated quasiparticle poles, the only singularity is a branch cut
in the imaginary axis $-iv_\mathrm{F}k<s<iv_\mathrm{F}k$, a
consequence of Landau damping.

Having understood the analytic structure of the Laplace transform,
we can now proceed to study the time evolution via the inverse
Laplace transform (\ref{antilap}) by closing the Bromwich contour
wrapping around the cut $-iv_\mathrm{F}k<s<iv_\mathrm{F}k$. We
obtain
\begin{equation}
\delta \theta_\bk(t)=
\int^1_{-1}\frac{d\alpha}{\alpha}\,\rho_\mathrm{ph}(\alpha)\cos(\alpha\tau)\,
\delta \theta_\bk(0), \label{realtimeev}
\end{equation}
where $\tau=v_\mathrm{F} k t$ and the spectral density for the
phase fluctuations is given by
\begin{equation}
\rho_\mathrm{ph}(\alpha)=\frac{1}{\pi}
\frac{f(x)F_\mathrm{I}(\alpha)}{\left[F_\mathrm{R}(\alpha)-f(x)\right]^2
+F_\mathrm{I}^2(\alpha)}. \label{specdens}
\end{equation}

\begin{figure}[t]
\includegraphics[width=3.5in,keepaspectratio=true]{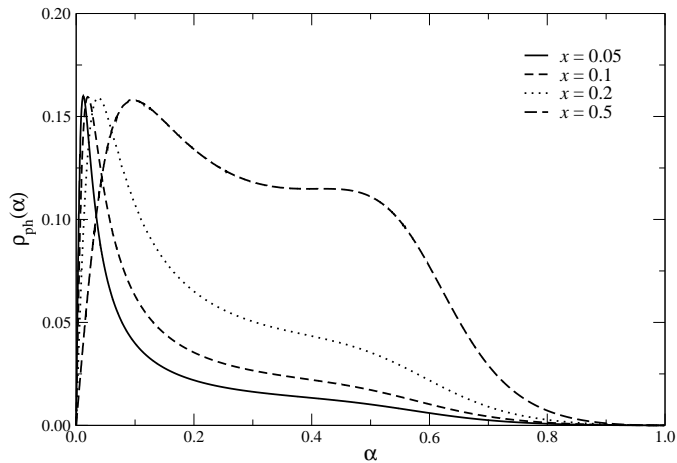}
\caption{Spectral density $\rho_\mathrm{ph}(\alpha)$ vs $\alpha$
in the Ginzburg-Landau regime.}\label{fig:rhoph1}
\end{figure}

\begin{figure}[t]
\includegraphics[width=3.5in,keepaspectratio=true]{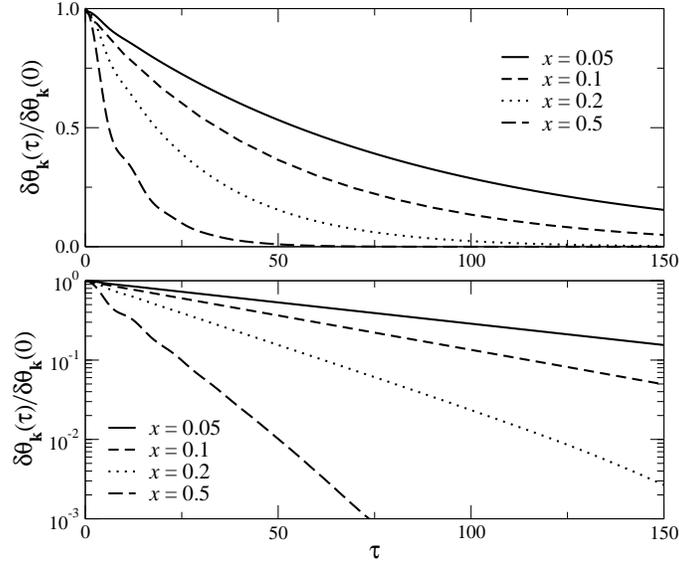}
\caption{Real-time evolution of the phase fluctuation
$\delta\theta_\bk(\tau)$ vs $\tau$ in the Ginzburg-Landau regime
is plotted in linear scale (top) and logarithmic scale
(bottom).}\label{fig:realtimeph1}
\end{figure}

The spectral density $\rho_\mathrm{ph}(\alpha)$ and the real-time
evolution of $\delta\theta_\bk(t)$ for several values of
$x\lesssim 1$ are displayed in Figs.~\ref{fig:rhoph1} and
\ref{fig:realtimeph1}, respectively. Three important features are
gleaned from these figures:
\begin{itemize}
\item[(i)]{The spectral density features a sharp peak at a value
$\alpha_\mathrm{peak}(x)$ that vanishes continuously as $x\to 0$.
This peak would indicate quasiparticle excitations with dispersion
relation $\omega=\alpha_\mathrm{peak}(x)v_\mathrm{F}k$. The group
velocity $v_\mathrm{g}(x)\simeq 14\,\zeta(3)v_\mathrm{F}x/3\pi^3$
vanishes at the critical point $x=0$ and increases linearly with
$x$ at least within the range $0<x\lesssim 0.2$.}
\item[(ii)]{While the spectral density is not of
the Breit-Wigner type (Lorentzian or resonance) and hence a true
width cannot be extracted unambiguously, it is clear from the
figure that qualitatively the quasiparticle excitations have a
``width''. This width vanishes at the critical point and increases
monotonically with $x \ll 1$ at least within the range consistent
with a Ginzburg-Landau expansion. Therefore these quasiparticles
will be Landau damped below $T_c$ and the relaxation time scale
(the inverse of the ``width'') diverges at the critical point
$x=0$. This expectation will be confirmed below and suggests
critical slowing down of long-wavelength phase fluctuations.}
\item[(iii)]{The reason that despite the appearance of a sharp
peak in the spectral density near the critical point the real-time
dynamics does not reveal the oscillations associated with a
``quasiparticle pole'' is clear. As $x\to 0$ \emph{both} the
damping rate and the group velocity $v_\mathrm{g}(x)$ vanish in
such a way that the time scale for damping is either shorter or of
the same order as the time scale for the oscillation.}
\end{itemize}
Furthermore, (\ref{realtimeev}) evaluated at $t=0$ leads to the
sum rule
\begin{equation}
\int^1_{-1}\frac{d\alpha}{\alpha}\,\rho_\mathrm{ph}(\alpha)
=1,\label{sumrule}
\end{equation}
which we have confirmed numerically for a wide range of $x$.

It is clear that damping becomes more pronounced for larger $x$
and while the peak would seem to lead to oscillations with period
$2\pi/\alpha_\mathrm{peak}(x)$ there are no hints of oscillatory
behavior in the real-time evolution. Phase fluctuations are
strongly overdamped without featuring a propagating mode. Hence we
conclude that near the critical point, in the Ginzburg-Landau
regime $x \ll 1 $, Goldstone modes or phase fluctuations are
severely damped despite the fact that the spectral density
features a peak that would indicate a quasiparticle ``dispersion
relation'' $\omega = \alpha_\mathrm{peak}(x) k$. The damping
becomes larger for larger $x$ and is solely a consequence of
\emph{collisionless} Landau damping.

For $x\ll 1$, we find that the nonequilibrium relaxation of the
phase fluctuation is very well approximated by an exponential (see
the logarithmic plot in Fig.~\ref{fig:realtimeph1}). The numerical
analysis clearly indicates that
$\delta\theta_\bk(t)\simeq\delta\theta_\bk(0)\,e^{-\gamma_k(x)t}$,
where from (\ref{IaJa}) and (\ref{realimag}) the damping rate
$\gamma_k(x)$ is found to be given by
\begin{equation}
\gamma_k(x)\simeq\frac{14\,\zeta(3)v_\mathrm{F} k}{3\pi^3}x
\quad\mathrm{for} \quad x\ll 1.\label{damprate}
\end{equation}
This result reveals \emph{critical slowing down} since the damping
rate vanishes at the critical point. Furthermore we also see that
for fixed $x$ the damping rate also vanishes in the
long-wavelength limit, in agreement with the expectation that the
relaxation time scale of Goldstone bosons should diverge in the
long-wavelength limit. This is one of the novel results of this
study: The long-wavelength phase fluctuations are
\emph{overdamped} by Landau damping in the Ginzburg-Landau regime
but the damping rate vanishes at the critical point indicating
critical slowing down.

In the Ginzburg-Landau regime and for long-wavelength,
low-frequency fluctuations the nonequilibrium retarded
Ginzburg-Landau effective action for phase fluctuations (Goldstone
modes) is given to lowest order by (we have set $J_\theta =0$)
\begin{equation}\label{phaseeffact}
S^\mathrm{GL}_\mathrm{ph}[\delta \theta] = \mathcal{N}(0)
\frac{|\Delta_0|}{2T_c}\int d^3k d\omega \left\{\delta
\theta_{-\bk}(-\omega)(v_\mathrm{F} k)^2
\left[F(k,\omega)-f(x)\right] \delta \theta_\bk(\omega)\right\},
\end{equation}
up to an additive constant,  where $f(x)$ and
$F(k,\omega)=F(\alpha=\omega/v_\mathrm{F} k)$ are given by
(\ref{IaJa}) and (\ref{realimag}), respectively. The imaginary
part of $F(\omega,k)$ originates in Landau damping. This
long-wavelength, low-frequency effective action leads to the
equations of motion for long-wavelength phase fluctuations in the
linearized approximation valid in the Ginzburg-Landau regime. Thus
it can be genuinely called the effective time dependent
Ginzburg-Landau effective action. It is nonlocal in time as a
consequence of Landau damping and describes real-time relaxation
which is completely overdamped. The damping rate vanishes at the
critical temperature thus signaling critical slowing down.

\subsubsection*{Away from the Ginzburg-Landau regime: $x \gtrsim 1$}

While we have focused in the Ginzburg-Landau regime $x\ll 1$, we
now study the region $x \gtrsim 1$ \emph{away} from the domain of
validity of the Ginzburg-Landau expansion mainly with the purpose
of comparing our results to those obtained in
Refs.~\onlinecite{aitchison97,aitchison95}.

\begin{figure}[t]
\includegraphics[width=3.5in,keepaspectratio=true]{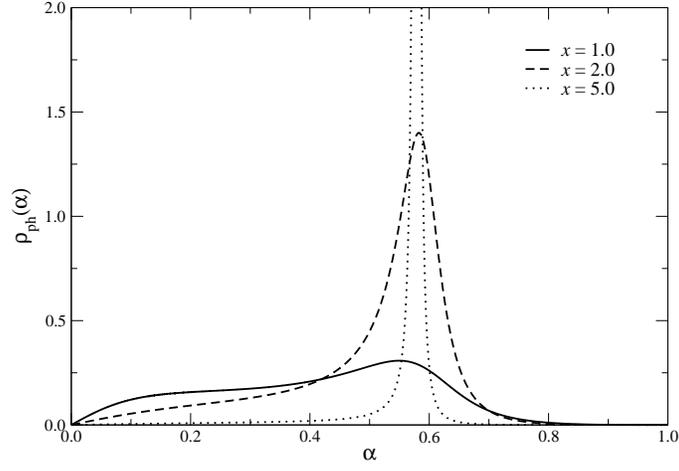}
\caption{Spectral density $\rho_\mathrm{ph}(\alpha)$ vs $\alpha$
away from the Ginzburg-Landau regime.}\label{fig:rhoph2}\figskip
\end{figure}

\begin{figure}[t]
\includegraphics[width=3.5in,keepaspectratio=true]{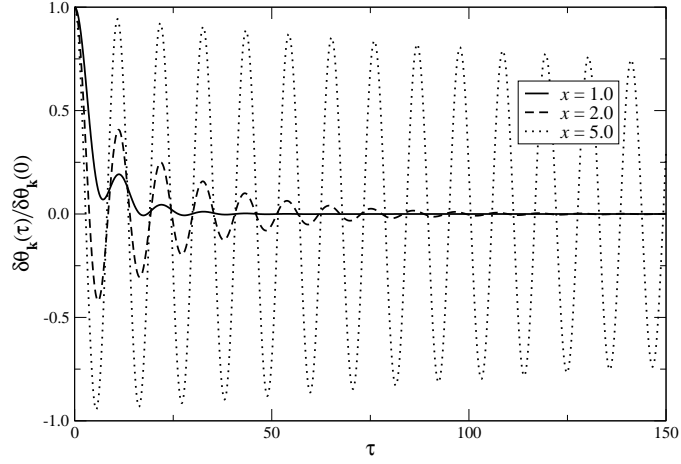}
\caption{Real-time evolution of the phase fluctuation
$\delta\theta_\mathbf{k}(\tau)$ vs $\tau$ away from the
Ginzburg-Landau regime.} \label{fig:realtimeph2}
\end{figure}

In this case we must keep the full $x$ dependence of the functions
$I_{a,b}$ and $J_{a,b}$. The spectral density has the same form as
in (\ref{specdens}) but now with the replacement $F(\alpha)\to
F(\sbar=-i\alpha+0^+,x)$, whose real [$F_\mathrm{R}(\alpha,x)$]
and imaginary [$F_\mathrm{I}(\alpha,x)$] parts can be found
straightforwardly as in the previous case. In the low-temperature
limit $x\gg 1$, we obtain
\begin{eqnarray}
f(x)&\simeq&\frac{1}{12x},\nn\\
F_\mathrm{R}(\alpha,x)&\simeq& \frac{\alpha^2}{4x},\nn\\
F_\mathrm{I}(\alpha,x)&\simeq&\frac{\pi\alpha}{8}
\bigg[\sqrt{1-\alpha^2}\,e^{-x/\sqrt{1-\alpha^2}}+\frac{\alpha^2
x}{2}\left[e^x\mathrm{Ei}(-x_+)+e^{-x}\mathrm{Ei}(-x_-)\right]\bigg],
\end{eqnarray}
where $\mathrm{Ei}(x)$ is the exponential integral function,
$x_\pm=x\left(\frac{1}{\sqrt{1-\alpha^2}}\pm 1\right)$ and the
exponentially small temperature corrections to $f(x)$ and
$F_\mathrm{R}(\alpha,x)$ have been neglected. The spectral density
with the full $x$ dependence is displayed in Fig.~\ref{fig:rhoph2}
for $x\gtrsim 1$ away from the regime of validity of the
Ginzburg-Landau approximation. It shows clearly the emergence of a
sharp quasiparticle peak, which for $x\gg 1$ is at
$\alpha_\mathrm{peak}=1/\sqrt{3}$ in agreement with the results of
Refs.~\onlinecite{aitchison97,aitchison95}. The real-time
evolution of phase fluctuations in this regime is displayed in
Fig.~\ref{fig:realtimeph2}. It is clear from these figures that
the sharp quasiparticle peak in the spectral density results in a
real-time dynamics that is weakly \emph{underdamped} by Landau
damping. This is in contrast to the real-time dynamics in the
Ginzburg-Landau regime, where Landau damping is so severe that the
phase fluctuation is overdamped and hence there is no
quasiparticle interpretation.

Thus the real-time evolution displayed above for this case
confirms the results of Refs.~\onlinecite{aitchison97,aitchison95}
valid well below the critical temperature, that a narrow
quasiparticle peak describes the dynamics of phase fluctuations.
As in the previous case discussed above, the damping rate is found
to be given by
\begin{equation}
\gamma_k(x)\simeq \frac{5 \pi v_\mathrm{F} k}{6} e^{-\sqrt{\frac{3}{2}}x}
\quad\mathrm{for}\quad x\gg 1, 
\end{equation}
where the factor $v_\mathrm{F} k$ is again a consequence of the
Goldstone nature of the phase fluctuation, leading to a vanishing
damping rate in the long-wavelength limit. Because the effects of
Landau damping are suppressed in the low-temperature limit ($x\gg
1$), the damping rate becomes smaller as the temperature decreases
and hence the oscillatory behavior associated with the
``quasiparticle pole'' is evidenced. In this region a \emph{local}
time-dependent effective theory is a good approximate description
of the nonequilibrium dynamics. On the contrary, the real-time
dynamics in the Ginzburg-Landau regime ($x\ll 1$) is
\emph{overdamped}, dominated by Landau damping and \emph{cannot}
be accurately described by a local effective action in real time.
Thus the Ginzburg-Landau dynamics is \emph{purely dissipative}.

\subsection{Amplitude fluctuations}

For the amplitude fluctuations in the long-wavelength,
low-frequency limit the equation of motion in terms of the Laplace
transform requires the  inverse propagator $1+g
\widetilde{\Sigma}_D(k,s)+g\widetilde{\Sigma}_O(k,s)$. In the case
of amplitude fluctuations we expect a ``gap'' of order
$|\Delta_0|$ in the spectrum of the quasiparticle excitations.
While this pole is away from the region of validity of the
long-wavelength, low-frequency approximation we can, nevertheless,
obtain a qualitative if not a quantitative estimate of the
dispersion relation for amplitude fluctuations.

\emph{Isolated poles}: Since the imaginary parts of $\Sigma_D$,
$\Sigma_O$ are nonzero only for
$-v_\mathrm{F}k<\omega<v_\mathrm{F}k$, for $k\simeq 0$ and
$\omega\simeq |\Delta_0|$ the expected quasiparticle pole will be
away from the continuum. We can find the position of this pole by
looking for the solutions of
\begin{equation}
\mathrm{Re}\left[1+g
\widetilde{\Sigma}_D(k,s)+g\widetilde{\Sigma}_O(k,s)
\right]_{s=-i\omega+0^+}=0.\label{poleamp}
\end{equation}
This equation has solutions for values of $\omega$ given by
$\omega_\mathrm{amp}(k)$, which determine the dispersion relation
for the amplitude fluctuation.

In the Ginzburg-Landau regime the gap of the spectrum can be
estimated by setting $k=0$ in the expressions of the self-energies
and keeping the lowest order [$\mathcal{O}(s^2)$] terms in the
expansion of the self-energies. After rescaling variables in the
integrals, we obtain
\begin{equation}
1+g\widetilde{\Sigma}_D(0,s)+g\widetilde{\Sigma}_O(0,s)\simeq
g\mathcal{N}(0)\int^{\infty}_{0}dz\frac{T(x,z)}{\epsilon^3}
\left[1+\frac{z^2 s^2}{4\epsilon^2|\Delta_0|^2}\right],
\end{equation}
where use has been made of the gap equation. Keeping the lowest
order in $x\ll 1$ in the Ginzburg-Landau regime, the integrals can
be done straightforwardly and we find
\begin{equation}
1+g\widetilde{\Sigma}_D(0,s)+g\widetilde{\Sigma}_O(0,s)\simeq
g\mathcal{N}(0)\frac{\pi
x}{4}\left[1+\frac{s^2}{8|\Delta_0|^2}\right],
\end{equation}
which, upon the analytic continuation $s\to -i\omega+0^+$,
suggests the gap of the spectrum in the Ginzburg-Landau regime to
be given by $2\sqrt{2}|\Delta_0|$.

\begin{figure}[t]
\includegraphics[width=3.5in,keepaspectratio=true]{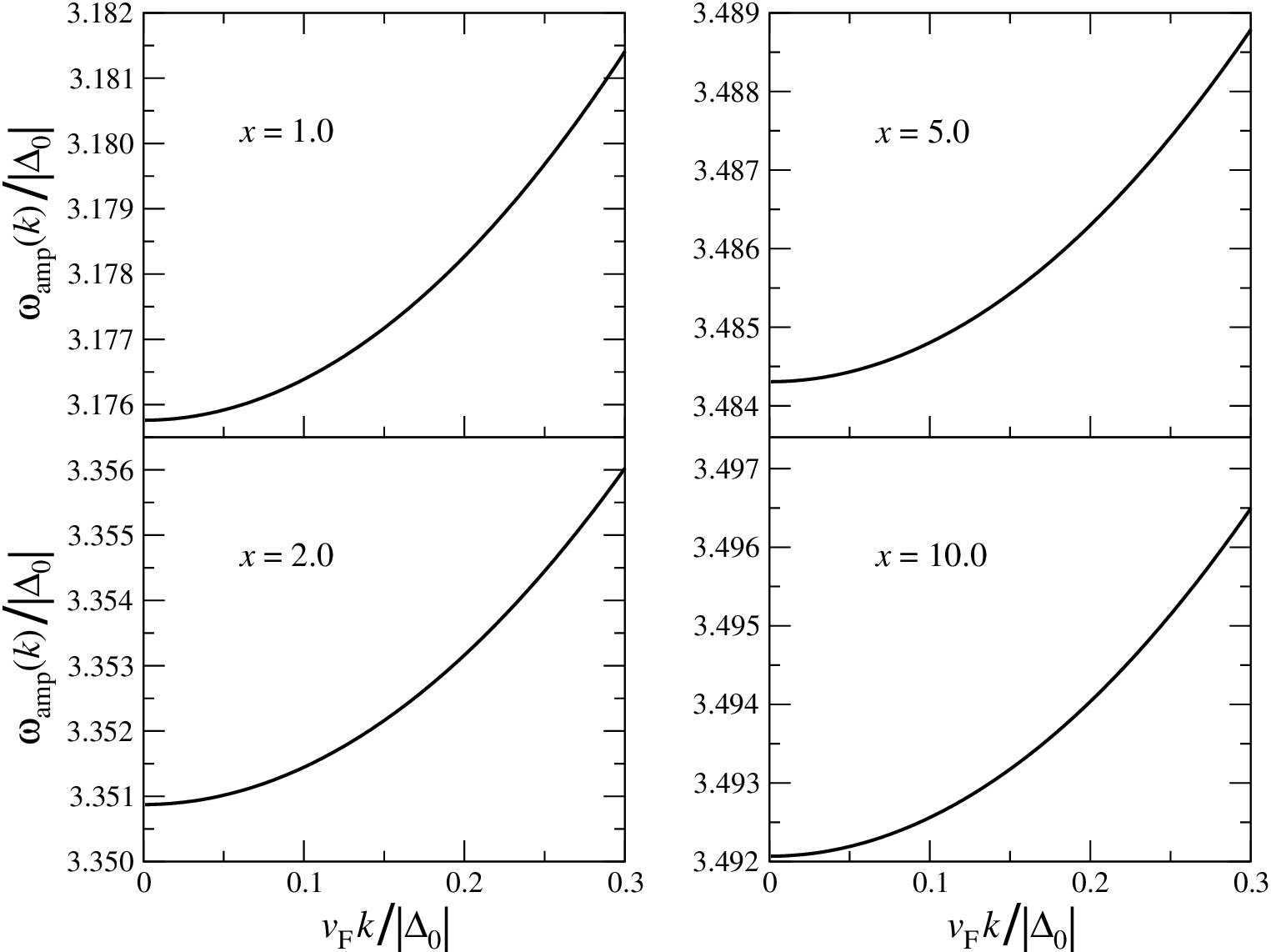}
\caption{Dispersion relation for the amplitude fluctuation
$\omega_\mathrm{amp}(k)$ away from the Ginzburg-Landau regime.}
\label{fig:amppoles}
\end{figure}

Away from the Ginzburg-Landau regime and for arbitrary values of
$k$ the equation for the dispersion relation, (\ref{poleamp}) must
be solved numerically. Figure~\ref{fig:amppoles} displays the
dispersion relation for several values of $x$ away from the
Ginzburg-Landau regime. The values of the gap shown are consistent
with the results obtained by Aitchison et al.
\cite{comparison2}. We haste to emphasize, however, that the
position of these single (quasi)particle poles are \emph{away}
from the regime of validity of our approximations and must only be
taken as indicative and consistent with the findings of
Ref.~\onlinecite{aitchison97} but \emph{not} as accurate
dispersion relations since higher orders in the ratio
$s/|\Delta_0|$ will modify these results.

This analysis is included here with the sole purposes of (i)
emphasizing that there are single (quasi)particle poles away from
the Landau damping continuum, consistent with the expectation of a
gap in the spectrum of small amplitude fluctuations, (ii)
establishing a comparison with the results of
Ref.~\onlinecite{aitchison97}, and (iii) offering at least a
qualitative, if not a reliable quantitative, discussion of the
terms contributing to the time evolution of amplitude
fluctuations. The full dispersion relations must be obtained by
keeping \emph{all} terms in the self-energies which will involve a
substantial numerical effort, a task clearly beyond the scope of
this article whose focus is on the Ginzburg-Landau regime in the
long-wavelength, low-frequency limit.

\emph{Landau damping cut}: The imaginary part associated with
Landau damping arises for $k\neq 0$ and is nonvanishing only along
the Landau damping cut $-v_\mathrm{F}k<\omega<v_\mathrm{F}k$.
Focusing on the long-wavelength, low-frequency limit and in the
Ginzburg-Landau regime, we find
\begin{eqnarray}
1+g\widetilde{\Sigma}_D(k,s)+g\widetilde{\Sigma}_O(k,s) &\simeq&
2g\left[\Sigma_O(0,0)+i\sbar x \mathcal{N}(0)
I_c(\sbar,x)\right],\label{ampcontinuum}
\end{eqnarray}
where we have used the Ward identity (\ref{WI}) and neglected the
contributions from $I_{a,b}$ and $J_{a,b}$, which are subdominant
for $v_\mathrm{F}k, s\ll|\Delta_0|$. We note that in the case of
phase fluctuations the terms with $I_c$ and $J_c$ cancel each
other in the difference in the self-energies because $I_c=J_c$,
however, for amplitude fluctuations these terms add up and furnish
the dominant contribution.

\begin{figure}[t]
\includegraphics[width=3.5in,keepaspectratio=true]{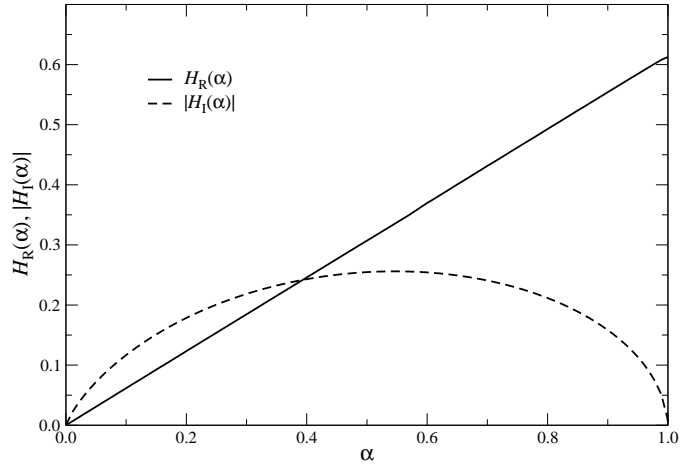}
\caption{Real and the imaginary parts of
$H(\alpha)$.}\label{fig:hofalfa}
\end{figure}

We now proceed to study the real-time evolution of the amplitude
fluctuation by integrating $\widetilde{\delta\rho}_\bk(s)$ along
the Bromwich contour in the complex $s$ plane parallel to the
imaginary axis and to the right of all the singularities of the
Laplace transform
\begin{equation}
\frac{\widetilde{\delta\rho}_\bk(s)}{{\delta\rho}_\bk(0)}=\frac{1}{s}
\frac{H(\sbar,x)}{H(\sbar,x)+h},\label{laplarho}
\end{equation}
where
\begin{equation}
xh=\Sigma_O(0,0)/\mathcal{N}(0),\quad H(\sbar,x)=i\sbar
I_c(\sbar,x).
\end{equation}
We obtain
\begin{equation}
\delta\rho_\bk(t)=\int^\infty_{-\infty}\frac{d\omega}{\omega}\,
\rho_\mathrm{amp}(\omega,k) \cos(\omega t)\delta\rho_\bk(0),
\label{ampoft}
\end{equation}
where $\rho_\mathrm{amp}(\omega,k)$ is the spectral density for
the amplitude fluctuation
\begin{eqnarray}
\rho_\mathrm{amp}(\omega,k)&=&\mathrm{sgn}(\omega)\,Z(k)\,
\omega^2_\mathrm{amp}(k)\,\delta[\omega^2-\omega^2_\mathrm{amp}(k)]
+\rho^\mathrm{cut}_\mathrm{amp}(\alpha),\nn\\
\rho^\mathrm{cut}_\mathrm{amp}(\alpha)&=&\frac{1}{\pi}\,
\mathrm{Im}\left[\frac{h}{h+H(\sbar,x)}\right]_{\sbar=-i\alpha+0^+}.
\label{ampspecdens}
\end{eqnarray}
In the above expression the first term arises from quasiparticle
pole $\omega_\mathrm{amp}(k)$ and the second term arises from the
Landau damping cut $-v_\mathrm{F}k<\omega<v_\mathrm{F}k$.

\begin{figure}[t]
\includegraphics[width=3.5in,keepaspectratio=true]{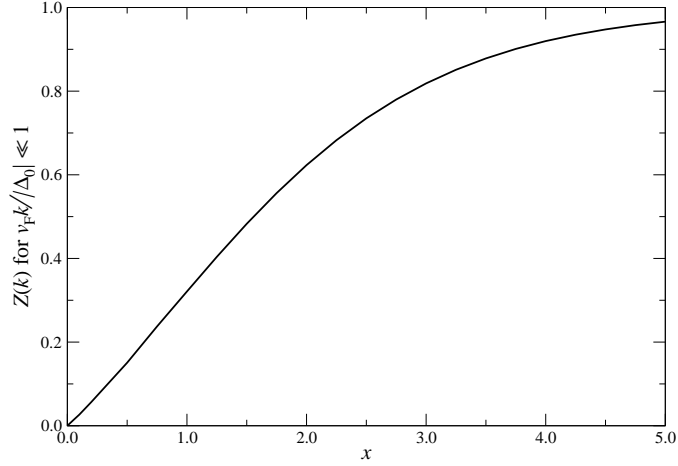}
\caption{Temperature dependence of the residue for the
quasiparticle pole $Z(k)$ in the long-wavelength
limit.}\label{fig:zamp}
\end{figure}

The quasiparticle pole will contribute an undamped oscillatory
component to the time evolution given by
\begin{equation}
\delta\rho^\mathrm{pole}_\bk(t)=Z(k)\cos[\omega_\mathrm{amp}(k)t].
\label{poletime}
\end{equation}
The residue of the quasiparticle pole $Z(k)$ is determined by
\begin{equation}
Z(k)=1-\int^1_{-1}\frac{d\alpha}{\alpha}\,
\rho^\mathrm{cut}_\mathrm{amp}(\alpha).
\end{equation}
Figure~\ref{fig:zamp} shows the temperature dependence of $Z(k)$
in the long-wavelength limit $v_\mathrm{F}k \ll |\Delta_0|$. It
reveals clearly that in the Ginzburg-Landau regime ($x\ll 1$) the
spectral density $\rho_\mathrm{amp}(k,\omega)$ is dominated by the
Landau damping cut.

Whereas the damping rate of the quasiparticle vanishes in the
long-wavelength, low-frequency approximation, we expect that
higher order contributions, in particular the decay into a pair of
Bogoliubov quasiparticles (bogolons) \cite{ketterson}, will lead
to a width of the quasiparticle pole and hence a finite damping
rate if the gap the quasiparticle spectrum is greater than
$2|\Delta_0|$, as seems to be the case from the previous analysis.
However, a more comprehensive study of the dispersion relation is
needed before reaching a quantitative conclusion. Again this is
beyond the regime of validity of the long-wavelength,
low-frequency approximation studied here.

\begin{figure}[t]
\includegraphics[width=3.5in,keepaspectratio=true]{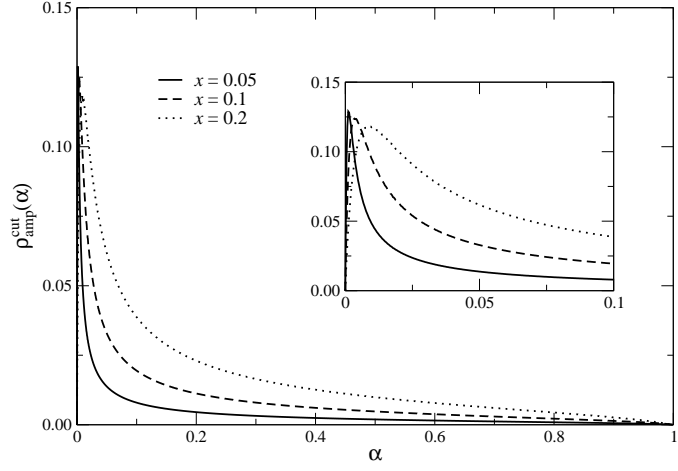}
\caption{Spectral density $\rho^\mathrm{cut}_\mathrm{amp}(\alpha)$
vs $\alpha$ in the Ginzburg-Landau
regime.}\label{fig:rhoamp}\figskip
\end{figure}

In the Ginzburg-Landau regime $x\ll 1$, we find [see
(\ref{LGcoefs})]
\begin{equation}
h=\frac{7\zeta(3)}{8\pi^2}x,
\end{equation}
and the real and imaginary parts of $H(\sbar=-i\alpha+0^+,0)$ to
be given by (here we have set $x=0$)
\begin{eqnarray}
H_\mathrm{R}(\alpha)&=& \frac{\alpha}{8} \int^{\infty}_0
\frac{dz}{z\epsilon}\ln\left|\frac{\alpha+
\frac{z}{\epsilon}}{\alpha-\frac{z}{\epsilon }}\right|,\nn\\
&\stackrel{\alpha\ll 1}\simeq&\frac{\pi^2|\alpha|}{16},\nn\\
H_\mathrm{I}(\alpha)&=& -\frac{\pi\alpha}{8}\Theta(1-\alpha^2)
\int_{z_\mathrm{min}}^\infty\frac{dz}{z\epsilon},\nn\\
&\stackrel{\alpha\ll
1}\simeq&-\frac{\pi\alpha}{8}\ln\frac{2}{\alpha}, \label{Hofalfa}
\end{eqnarray}
respectively, where $z_\mathrm{min}=\sqrt{\alpha^2/(1-\alpha^2)}$.
The real and imaginary parts of $H(\alpha)$ are displayed in
Fig.~\ref{fig:hofalfa}. The Landau damping contribution to the
spectral density in the Ginzburg-Landau regime and the
long-wavelength, low-frequency limit is accurately described by
\begin{equation}
\rho^\mathrm{cut}_\mathrm{amp}(\alpha)=
\frac{h\,|H_\mathrm{I}(\alpha)|}
{\big[h+H_\mathrm{R}(\alpha)\big]^2
+H^2_\mathrm{I}(\alpha)}.\label{ampsp}
\end{equation}
which, as displayed in Fig.~\ref{fig:rhoamp}, clearly reveals a
sharp peak near $\alpha\approx 0$.

\begin{figure}[t]
\includegraphics[width=3.5in,keepaspectratio=true]{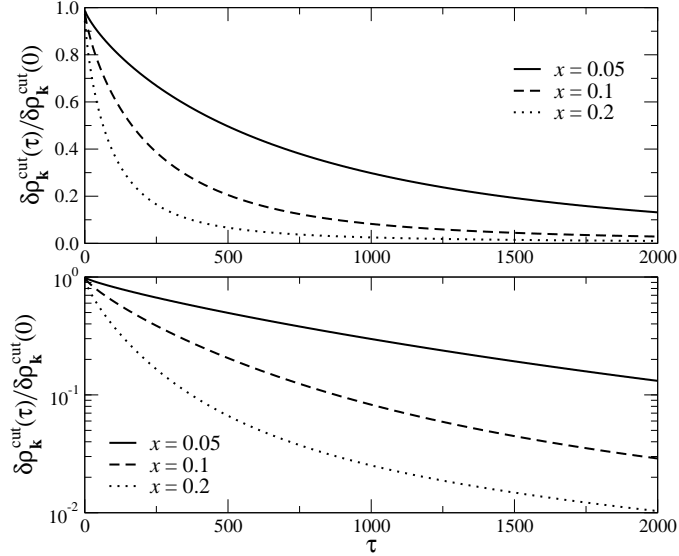}
\caption{Real-time evolution of the Landau damping contribution to
the amplitude fluctuation $\delta\rho^\mathrm{cut}_\bk(t)$ in the
Ginzburg-Landau regime is plotted in linear scale (top) and
logarithmic scale (bottom).}\label{fig:realtimeamp}
\end{figure}

The real-time evolution of the Landau damping contribution to the
amplitude fluctuation $\delta\rho^\mathrm{cut}_\bk(t)$ is
displayed in Fig.~\ref{fig:realtimeamp}. While for $x\lesssim
0.05$ the nonequilibrium relaxation through Landau damping can be
approximated by an exponential, clearly such is not the case for
$x\gtrsim 0.05$. Nevertheless we still find that the Landau
damping relaxation time scale becomes \emph{longer} as $x\to 0$,
revealing that this contribution to the nonequilibrium relaxation
seems to be critically slowed down near the critical point.

We close this section, by providing the nonlocal Landau damping
contribution to the \emph{retarded} effective action for
long-wavelength, low-frequency amplitude fluctuations in the
Ginzburg-Landau regime (we have set $J_\rho =0$)
\begin{equation}
S^\mathrm{GL}_\mathrm{amp}[\delta\rho]= 2
\mathcal{N}(0)\frac{|\Delta_0|}{T_c}\int d^3k d\omega\,
\delta\rho_{-\bk}(-\omega)\left[\frac{7\zeta(3)|\Delta_0|}{8\pi^2T}
+H(\omega,k)\right]\delta\rho_\bk(\omega),\label{ampeffac}
\end{equation}
up to a additive constant, where
$H(\omega,k)=H(\alpha=\omega/v_\mathrm{F}k)$ given by
(\ref{Hofalfa}). Obviously,
$S^\mathrm{GL}_\mathrm{amp}[\delta\rho]$ leads to the retarded
equations of motion for the long-wavelength, low-frequency
amplitude fluctuations.

\section{Conclusions}\label{sec:conclusions}

In this article we have focused on the real-time nonequilibrium
dynamics of small phase and amplitude fluctuations of the order
parameter in neutral BCS superconductors in the Ginzburg-Landau
regime near the critical point. We have implemented the
Schwinger-Keldysh formulation of nonequilibrium field theory
combined with the novel tadpole method to obtain directly in real
time the \emph{retarded} equations of motion for small
fluctuations around equilibrium. These equations allow to extract
the one-loop effective action for the long-wavelength,
low-frequency phase and amplitude fluctuations in the
Ginzburg-Landau regime, which is characterized by $|\Delta_0(T)|/T
\ll 1$ with $\Delta_0(T)$ the finite-temperature gap. Furthermore,
the retarded equations of motion can be cast as an initial value
problem to study the relaxation of nonequilibrium fluctuations
directly in real time.

We studied in detail the relaxation of phase fluctuations within
and away from the Ginzburg-Landau regime. Despite the fact that
the spectral density features a sharp peak with a Goldstone-like
dispersion relation in the Ginzburg-Landau regime the relaxation
is completely \emph{overdamped} as a consequence of Landau
damping. This is consistent with a purely dissipative
time-dependent  Ginzburg-Landau equation. However, the effective
action is nonlocal because of Landau damping. The relaxation is
exponential in time with the damping rate
$\gamma_k(T)=14\zeta(3)v_\mathrm{F}k|\Delta_0(T)|/3\pi^3T$. The
factor $v_\mathrm{F} k$ is a consequence of the Goldstone nature
of the phase fluctuations. The relaxation of phase fluctuations
near the critical point features \emph{critical slowing down},
i.e., the relaxation time scale diverges at the critical point.

Far away from the Ginzburg-Landau regime at low temperatures, the
spectral density features sharp quasiparticle peaks and the
nonequilibrium relaxation is \emph{underdamped} in agreement with
the results of Refs.~\onlinecite{aitchison95,aitchison97}. Away
from the critical region, the contribution from Landau damping is
negligible.  The long-wavelength amplitude fluctuations are
severely Landau-damped near the critical region and the relaxation
time scales also feature critical slowing down.

While we have focused on the nonequilibrium dynamics of neutral
BCS superconductors, as a next step we will apply these methods to
the case of charged superconductors to study in detail the
dynamics of the Carlson-Goldman modes as well as the effective
action including gauge fields near the critical temperature. We
postpone this study to a forthcoming article.

\begin{acknowledgments}
We thank I.\ J.\ R.\ Aitchison and V.\ P.\ Gusynin for helpful and
illuminating correspondence. S.M.A.\ would like to thank King Fahd
University of Petroleum and Minerals for financial support. The
work of D.B.\ was supported in part by the US NSF under grant
PHY-9988720. The work of S.-Y.W.\ was supported in part by the US
DOE under contract W-7405-ENG-36.
\end{acknowledgments}

\appendix

\section{Plane wave solutions and the Bogoliubov transformation}\label{app:BT}

In this Appendix we present an alternative derivation of the
correlation functions for the fields $\psi_\sigma$,
$\psi_\sigma^\dagger$ directly from the plane wave solutions of
the homogeneous equations of motion (i.e., in the absence of
source) for the Nambu-Gor'kov field given by (\ref{eqnofmot})
\begin{equation}
\left[i \frac{\partial}{\partial t}+\sigma_3\left(
\frac{\nabla^2}{2m}+\mu\right)+\sigma_+ \Delta_0 +
\sigma_-\Delta_0^\ast\right]\Psi(\xt)= 0.\label{heqnofmot}
\end{equation}
The plane wave solution can be written in the form
\begin{equation}
\Psi(\bx,t) = \Phi(k)\,e^{-i(\omega t- \bk \cdot \bx)},\quad
\Phi(k)=
\begin{bmatrix}
U_k \\V_k
\end{bmatrix}. \label{spinorsol}
\end{equation}
The two-component Nambu-Gor'kov spinor obeys
\begin{equation}
\begin{bmatrix}
\xi_k & -\Delta_0 \\
-\Delta_0^\ast & -\xi_k
\end{bmatrix}
\begin{bmatrix}
U_k \\V_k
\end{bmatrix} = \omega \begin{bmatrix}
U_k \\V_k
\end{bmatrix}, \label{matxeqn}
\end{equation}
where $\xi_k = k^2/2m-\mu$. This is an eigenvalue equation with
the eigenvalues given by $\omega= \pm E_k$, where
$E_k=\sqrt{\xi^2_k+|\Delta_0|^2}$. The normalization of the
positive and negative energy spinors is chosen so that
$\Phi^{(\alpha)\dagger}(k)\Phi^{(\beta)}(k) =
\delta_{\alpha\beta}$, where $\alpha,\beta=1,2$ (not to be
confused with the Nambu-Gor'kov indices) correspond to $\omega=\pm
E_k$, respectively. Introducing the Bogoliubov coefficients $u_k$,
$v_k$ given by
\begin{equation}
u_k= \left(\frac{\xi_k+E_k}{2E_k} \right)^{1/2}, \quad
v_k=u_k\left(\frac{\Delta_0^\ast}{\xi_k+E_k}\right),\quad u_k
v_k=\frac{\Delta_0^\ast}{2E_k},\label{coefs}
\end{equation}
and satisfying $u_k^2+|v_k|^2=1$, we find that the positive and
negative energy spinors are given by
\begin{gather}
\Phi^{(1)}(k) = \begin{bmatrix} u_k \\-v_k
\end{bmatrix},\quad
\Phi^{(2)}(k) = \begin{bmatrix}
v^\ast_k \\
u_k
\end{bmatrix},\label{negenspin}
\end{gather}
respectively. After accounting for the interpretation of negative
energy solutions as antiparticles, one can therefore write the
general plane wave solution of the homogeneous equation of motion
(\ref{heqnofmot}) as
\begin{equation}
\Psi(\bx,t) = \frac{1}{\sqrt{V}} \sum_{\bk} \left\{b_\bk
\begin{bmatrix}
u_k \\-v_k
\end{bmatrix}e^{-i(E_k t-\bk\cdot\bx)} +
d^\dagger_\bk
\begin{bmatrix}
v^\ast_k \\u_k
\end{bmatrix}
e^{i(E_k t-\bk\cdot\bx)}\right\},\label{spinsolu}
\end{equation}
where $V$ is the quantization volume.

At the level of second quantization, one recognizes that
\eqref{spinsolu} is the Bogoliubov transformation, in which the
operator $b^\dagger_\bk$, $d^\dagger_\bk$ create Bogoliubov
quasiparticles of momentum $\bk$ (energy $E_k$) and obey the usual
canonical anticommutation relations. The correlation functions of
the Nambu-Gor'kov fields in the density matrix that describes free
Bogoliubov quasiparticles in thermal equilibrium at inverse
temperature $\beta$ are therefore found to be given by (in the
continuum limit)
\begin{gather*}
\langle \Psi_a(\bx,t) \Psi^{\dagger}_b(\bx',t') \rangle =
\intk\left[[1-n_\mathrm{F}(E_k)]\,\mathcal{S}_{ab}(k)\,e^{-i
E_k(t-t')}+
n_\mathrm{F}(E_k)\,\overline{\mathcal{S}}_{ab}(k)\,e^{i
E_k(t-t')}\right]e^{i\bk\cdot(\bx-\bx')},\nn\\
\langle \Psi^{\dagger}_b(\bx',t') \Psi_a(\bx,t)\rangle =
\intk\left[n_\mathrm{F}(E_k)\,\mathcal{S}_{ab}(k)\,e^{-i
E_k(t-t')}+[1-n_\mathrm{F}(E_k)]\,\overline{\mathcal{S}}_{ab}(k)\,e^{i
E_k(t-t')}\right]e^{i\bk \cdot (\bx-\bx')},
\end{gather*}
where $n_\mathrm{F}(E_k)$ is the equilibrium distribution for
Bogoliubov quasiparticles of momentum $\bk$
\begin{equation}
n_\mathrm{F}(E_k)=\langle b^\dagger_\bk b_\bk\rangle=\langle
d^\dagger_\bk d_\bk\rangle=\frac{1}{e^{\beta E_k}+1},
\end{equation}
and $\mathcal{S}(k)$, $\overline{\mathcal{S}}(k)$ are given by
(\ref{gmatrices}).


\begin{thebibliography}{99}
\bibitem{comment}
For simplicity of presentation, we consider here only the
generating functional for the Nambu-Gor'kov fields $\Psi$ and
$\Psi^\dagger$, whose Green's functions will enter explicitly in
the equations of motion for the condenstate perturbation. The
inclusion of the (bosonic) pair fields $\chi$ and $\chi^\dagger$
in the generating functional is straightforward.
\bibitem{comparison2}
In the first reference in Ref.~\onlinecite{aitchison97}, it is
found that the value of the gap is $\approx 1/\sqrt{0.085} \simeq
3.43$ [see Eq.~(31) in that reference].
\end{thebibliography}
\end{document}